\documentclass[%
 reprint,
superscriptaddress,
 amsmath,amssymb,
]{revtex4-1}

\usepackage{bbold}
\usepackage{mathptmx}
\usepackage{subfig}
\usepackage{psfrag,graphicx}
\usepackage{dcolumn}
\usepackage{amsmath,amssymb}
\usepackage{bm}
\usepackage{color}
\usepackage{latexsym}
\usepackage{epstopdf}
\usepackage{color}
\usepackage[english]{babel}
\usepackage{latexsym}
\usepackage{psfrag,graphicx}
\usepackage{amsmath}
\usepackage{amssymb}
\usepackage{amsfonts}
\usepackage{bm}
\usepackage{natbib}
\usepackage{epstopdf}
\usepackage{appendix}
\usepackage{rotating}
\usepackage[english]{babel}
\usepackage{aeguill}
\usepackage{ulem}
\usepackage[justification=justified]{caption}

\definecolor{mygrey}{gray}{0.35}
\definecolor{myblue}{rgb}{0.2,0.2,0.8}
\definecolor{myzard}{cmyk}{0,0,0.05,0}
\definecolor{mywhite}{rgb}{1,1,1}
\definecolor{mywhite}{rgb}{1,1,1}
\definecolor{myred}{rgb}{1,0.,0.3}

\usepackage[colorlinks=true,citecolor=myblue,linkcolor=myblue]{hyperref}

\def\ba{\begin{align}}
\def\enda{\end{align}}
\def\bi{\begin{itemize}}
\def\ei{\end{itemize}}

\def\be{\begin{equation}}
\def\ee{\end{equation}}
\def\bea{\begin{eqnarray}}
\def\eea{\end{eqnarray}}
\def\bse{\begin{subequations}}
\def\ese{\end{subequations}}

\newcommand{\ket}[1]{|{#1}\rangle}                       
\newcommand{\average}[1]{\langle {#1} \rangle}           

\newcommand{\Ignore}[1]{ }

\def\i{\text{i}}

\begin{document}

\preprint{APS/123-QED}

\title{Quantum phase transitions for an integrable quantum Rabi-like model with two interacting qubits}

\author{R. Grimaudo}
\address{Department of Physics and Chemistry ``Emilio Segr\`{e}",
University of Palermo, viale delle Scienze, Ed. 18, I-90128, Palermo, Italy}

\author{A. S. Magalh{\~a}es de Castro}
\address{Universidade Estadual de Ponta Grossa, Departamento de F\'{\i}sica, CEP 84030-900, Ponta Grossa, PR, Brazil}

\author{A. Messina}
\address{ Department of Mathematics and Informatics, University of Palermo, Via Archirafi 34, I-90123 Palermo, Italy}

\author{E. Solano}
\address{International Center of Quantum Artificial Intelligence for Science and Technology (QuArtist) and
Physics Department, Shanghai University, 200444 Shanghai, China}
\address{Department of Physical Chemistry, University of the Basque Country UPV/EHU, Apartado 644, 48080 Bilbao, Spain}
\address{IKERBASQUE, Basque Foundation for Science, Plaza Euskadi 5, 48009 Bilbao, Spain}
\address{Kipu Quantum, Kurwenalstrasse 1, 80804 Munich, Germany}

\author{D. Valenti}
\address{Department of Physics and Chemistry ``Emilio Segr\`{e}",
University of Palermo, viale delle Scienze, Ed. 18, I-90128, Palermo, Italy}

\date{\today}

\begin{abstract}
 
A two-interacting-qubit quantum Rabi-like model with vanishing transverse fields on the qubit-pair is studied. Independently of the coupling regime, this model can be exactly and unitarily reduced to two independent single-spin quantum Rabi models, where the spin-spin coupling plays the role of the transverse field.
This transformation and the analytical treatment of the single-spin quantum Rabi model provide the key to prove the integrability of our model.
The existence of different first-order quantum phase transitions, characterized by discontinuous two-spin magnetization, mean photon number and concurrence, is brought to light.

\end{abstract}



\maketitle

\textit{Introduction.}
The Quantum Rabi Model (QRM) \cite{Rabi36, Rabi37, JC, Braak16} describes the simplest nontrivial coupling mechanism between a single qubit and a quantized bosonic mode.
In spite of an apparently simple bilinear coupling between fermionic and bosonic degrees of freedom, the complete list of its eigevalues has  been unveiled only in 2001 by Braak \cite{Braak11}.
The presence of counterrotating terms breaks the U(1)-symmetry of the rotating ones.
However, the remaining $Z(2)$-symmetry gives rise to a spectrum characterized by a complex structure. 

It has been recently demonstrated that the QRM exhibits a quantum phase transition (QPT) driven by the qubit-mode coupling \cite{Ashhab13, Wang15, Ying15, Ying22, Ying22new}.
The study of the QPTs is one of the most focused topics
in light-matter interaction systems \cite{Vojta05, Carollo20, Rossini21}, and obtaining analytical results is a difficult goal.

In many applications, e.g. in quantum computing \cite{Barenco95, Hua14}, it is necessary to consider more complex scenarios to perform controlled gates \cite{Romero12, Barends19} and to generate multipartite entangled states \cite{Kang16, Lu13, Li09}.
For such purposes, the generalized versions of the QRM are useful, for instance, the Dicke model \cite{Dicke54}, two-photon QRM \cite{Chen12, Felicetti15}, multi-photon QRM \cite{Zhang13}, multi-level QRM \cite{Albert12}, two-qubit QRM \cite{Agarwal12, Peng12, Lee13, Chilingaryan13, Wang14, Peng14}, multi-qubit QRM \cite{Peng21, Zhang21prl}.
The interest towards these models results to be of significant importance for circuit quantum electrodynamics \cite{Nataf11, Lizuain10} and semiconductor systems \cite{Crusotto13, Anappara09, Todorov10}.

In this work, we study a class of two-qubit QRMs where, besides the qubit-mode interaction, a qubit-qubit coupling is taken into account. Further, a longitudinal field is applied to the qubit pair, whereas a transverse field is absent. Such a model turns out to be profoundly different from the usual two-qubit QRM commonly analysed \cite{Zhang15, Duan15, Dong16, Mao19, Sun20, Yan21, Zhang21, Liu21, Mao21}, where a qubit-qubit interaction is missing. However, in some contexts such an interaction cannot be neglected, and in some scenarios, such as in quantum computation, is fundamental to perform two-qubit quantum logic gates to generate entangled states of the system \cite{Kang16, Lu13, Li09}.

Generally speaking, the determination of both eigenvalues and eigenvectors of a time-independent Hamiltonian is the basic necessary step for unveiling intriguing physical properties of the system.
Exact or approximate solutions of the standard two-qubit QRM have been obtained through the application of several methods, like Bargmann-space techniques \cite{Peng12,Peng14}, perturbation theory \cite{Chilingaryan13}, the generalized rotating-wave approximation (GRWA) \cite{Zhang15}, the method of extended coherent states \cite{Duan15}, adiabatic approximation and zeroth-order approximation method \cite{Mao15}.

We will show that, thanks to the existence of a constant of motion \cite{GVM1, GMNSV}, our model can be exactly, unitarily reduced to two independent, asymmetric \cite{Liu21} single-qubit QRMs. Here, the role of the (effective) transverse field is played by the qubit-qubit coupling. In this way, the model turns out to be integrable on the basis of the exact Braak's solutions of the single-qubit QRM \cite{Braak11}. A similar model has been analysed, with a different method, in Ref. \cite{Peng14}. Nevertheless, that model results to be non-integrable and no QPTs are present. In this work, instead, both the model and the used approach allow to easily and exactly identify the occurrence of QPTs, characterized by an abrupt change of the two-qubit magnetization, the mean photon number and the two-qubit level of entanglement. Such QPTs occur for a variation of both the qubit-mode coupling, the qubit-qubit-coupling, and the strength of the longitudinal magnetic field.

\textit{Model.}
Consider the following model (in units of $\hbar$):
\begin{equation} \label{Hamiltonian}
\begin{aligned}
{H} = &
{\varepsilon_1}\hat{\sigma}_{1}^{z}+{\varepsilon_2}\hat{\sigma}_{2}^{z} + \omega \hat{a}^\dagger \hat{a} \\
& + {\gamma_{x}}\hat{\sigma}_{1}^{x}\hat{\sigma}_{2}^{x} +
{\gamma_{y}}\hat{\sigma}_{1}^{y}\hat{\sigma}_{2}^{y}+
\gamma_{z}\hat{\sigma}_{1}^{z}\hat{\sigma}_{2}^{z} \\
& + (g_1 \hat{\sigma}_1^z + g_2 \hat{\sigma}_2^z) (\hat{a} +\hat{a}^\dagger) ,
\end{aligned}
\end{equation}
which describes two interacting spin-1/2's subject to local longitudinal ($z$) fields and coupled to the same single field mode through different (real) coupling parameters.
$\omega$ and $\epsilon_i$ ($i=1,2$) are the characteristic frequencies of the mode and the $i$-th spin, respectively.
$\hat{\sigma}_{k}^{l}$ ($k=1,2$, $l=x,y,z$) are the Pauli operators of the spins, while $(a,a^\dagger)$ are the annihilation and creation boson operators of the field mode.

Thanks to the existence of the constant of motion $\hat{\sigma}_{1}^{z}\hat{\sigma}_{2}^{z}$, the model can be unitarily transformed into $H = H_a \oplus H_b$, with
\begin{equation} \label{Ha Hb}
\begin{aligned}
{H}_{a} = &
{\varepsilon_a} \hat{\sigma}_{a}^{z} +
{\gamma_{a}} \hat{\sigma}_{a}^{x} + \gamma_{z} \hat{\mathbb{1}}_{a} +
\omega ~ \hat{a}^\dagger \hat{a} +
g_a \left( \hat{a}^\dagger + \hat{a} \right)  \hat{\sigma}_{a}^z, \\
{H}_{b} = &
{\varepsilon_b} \hat{\sigma}_{b}^{z} +
{\gamma_{b}} \hat{\sigma}_{b}^{x} - \gamma_{z} \hat{\mathbb{1}}_{b} +
\omega ~ \hat{a}^\dagger \hat{a} +
g_b \left( \hat{a}^\dagger + \hat{a} \right)  \hat{\sigma}_{b}^z, 
\end{aligned}
\end{equation}
where $\epsilon_{a/b} = \varepsilon_1 \pm \varepsilon_2$, $\gamma_{a/b} = \gamma_{x} \mp \gamma_{y}$ and $g_{a/b}=g_1 \pm g_2$.
The effective Hamiltonian $H_a$ ($H_b$) governs the dynamics of the two-spin-mode system within the dynamically invariant subspace $\mathcal{H}_a$ ($\mathcal{H}_b$) spanned by $\{ \ket{++},\ket{--} \} \otimes \{ \ket{n} \}_{n \in 0}^\infty$ ($\{ \ket{+-},\ket{-+} \} \otimes \{ \ket{n} \}_{n \in 0}^\infty$), with $\hat{\sigma}^z\ket{\pm}=\pm\ket{\pm}$ and $\hat{a}^\dagger \hat{a}\ket{n}=n\ket{n}$ (see supplemental material).
The two Hamiltonians in Eq. \eqref{Ha Hb} look like the well known asymmetric QRM \cite{Liu21}
\footnote{the original QRM is obtained by putting the longitudinal ($\hat{z}$) field equal to zero.
Moreover, a $\pi/2$-rotation around the $\hat{y}$-axis has to be performed for both $H_a$ and $H_b$ to get the standard form of the asymmetric QRM.}.
It is worth noticing that the role of the transverse field in the two effective Hamiltonians is played by the (effective) spin-spin couplings $\gamma_a$ and $\gamma_b$.

We emphasize that the two qubits behave as effective two-level systems within each invariant subspace.
Therefore, each information at our disposal or obtained for the effective dynamics of the \textit{fictitious} two-level systems described by $\sigma_a^l$ and $\sigma_b^l$ ($l=x,y,z$) can be reinterpreted in terms of the \textit{actual} two coupled qubits, $\sigma_1^l$ and $\sigma_2^l$, through the following mapping
\begin{equation} \label{Mapping}
    \ket{\pm}_a \Longleftrightarrow \ket{\pm\pm},
    \qquad
    \ket{\pm}_b \Longleftrightarrow \ket{\pm\mp},
\end{equation}
where $\ket{\pm}_a$ ($\ket{\pm}_b$) are the two single-spin states of the fictitious spin-$a$ (spin-$b$).
This means that the study of the dynamics of the original system (two interacting qubits coupled to the same field mode) can be reduced to that of two independent effective single-spin quantum Rabi problems.
In other words, we can solve the original dynamical problem by applying to each two-dimensional subdynamics the known results reported in literature for the single-spin QRM.
It is worth noticing that such a reduction based on an analytical method (see the supplemental material) is independent of the Hamiltonian parameters as well as of their possible time-dependence.
In particular, no constrains related to the spin-mode couplings are present: our approach holds for weak, strong, ultra-strong and deep-strong spin-mode coupling (see Refs. \cite{Kockum19, Xie17} for the classification of the coupling regimes).
Moreover, it is worth noting that when $g_1 \approx g_2$ the subspace $a$ can be characterized by either a weak or strong coupling regime (depending on the magnitude of the two couplings), while the subspace $b$ would ever be in the weak coupling regime.

\textit{Spectrum.}
Thanks to the exact dynamical decomposition which breaks down the initial dynamics into two independent easier (sub)dynamics, the eigenvalue problem can be successfully dealt with.
The spectrum of the two-spin-mode system, indeed, is obtained by the `union' of the spectra of $H_a$ and $H_b$, each of which can be analytically derived from the QRM spectrum \cite{Braak11} (see the supplemental material).
The latter presents two series of eigenvalues related to the different value of the parity \cite{Braak11}, and then our model is characterized, in general, by four series of eigenvalues.
The exact dynamical reduction of our two-qubit QRM implies the integrability of the model as direct consequence of that demonstrated by Braak \cite{Braak11} for the single-qubit QRM.

\begin{figure}[] 
\begin{center}
{\includegraphics[width=0.23\textwidth]{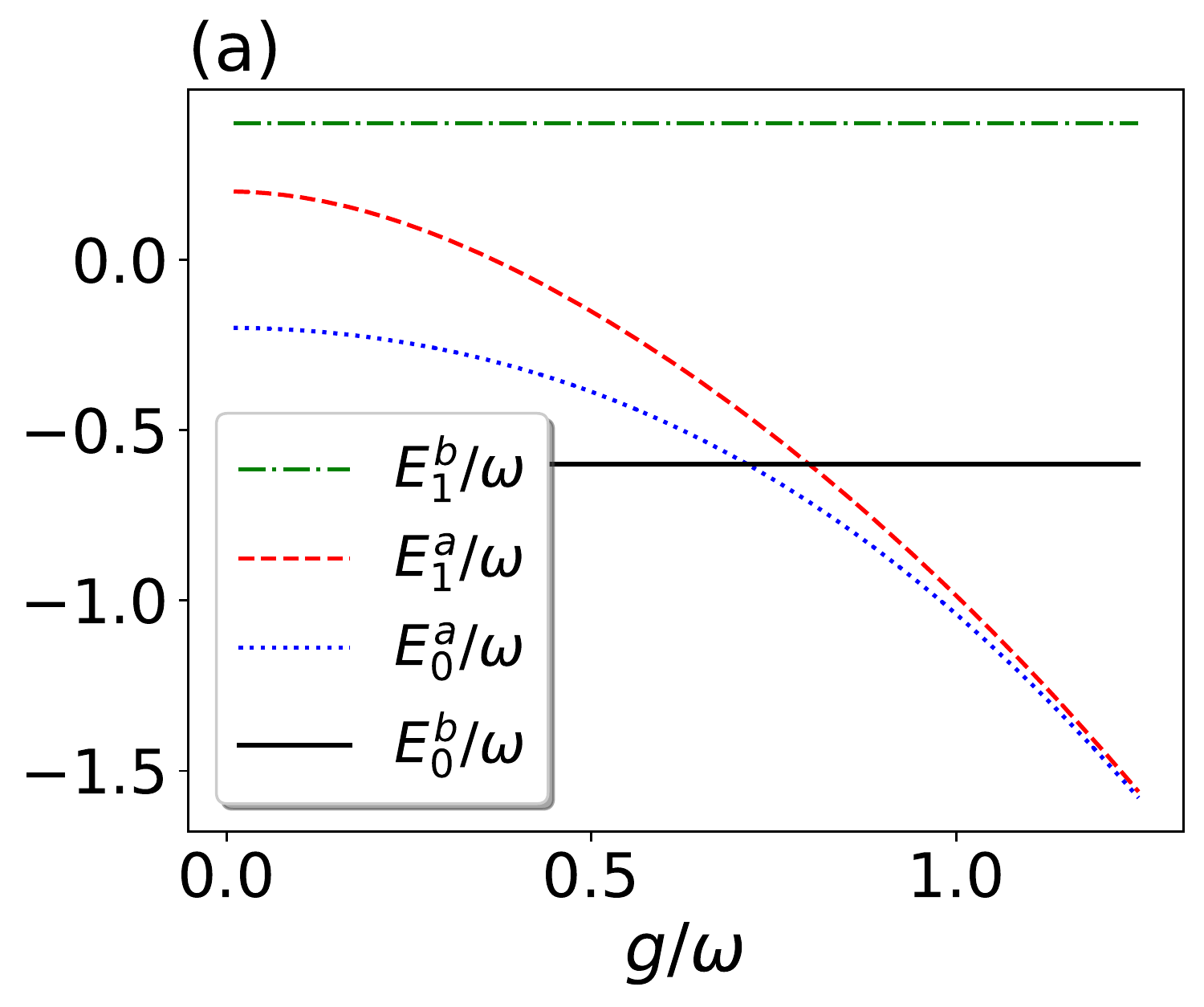}}
{\includegraphics[width=0.23\textwidth]{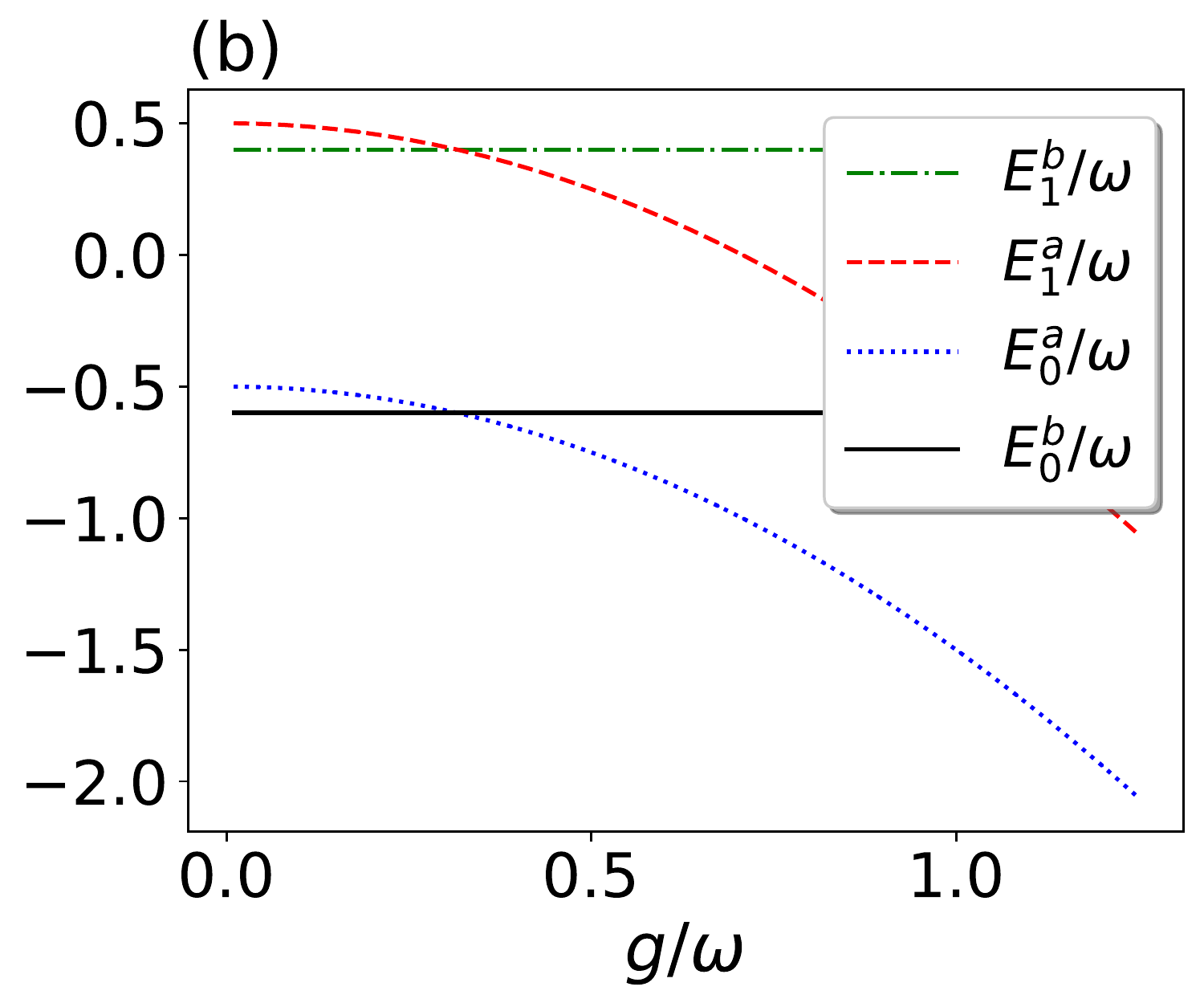}}
\captionsetup{justification=raggedright,format=plain,skip=4pt}%
\caption{First four eigenvalues of the two-spin-mode system for $g_1=g_2=g/2$ and: (a) $\varepsilon_1=\varepsilon_2=0$, $\gamma_x=2\gamma_y=0.4\omega$; (b) $\varepsilon_1=\varepsilon_2=0.25\omega$, $\gamma_x=\gamma_y=0.3\omega$.}
\label{fig: spectrum}
\end{center}
\end{figure}

Of course, depending on the Hamiltonian parameters, different scenarios with different eigenspectra can arise.
From now on, without loss of generality, the case $\gamma_z=0$ is considered.
Such a term only causes a shifting in the spectra of $H_a$ and $H_b$ with no relevant physical implications.
To appreciate this claim, let us consider the two special cases for which some terms of $H_a$ and/or $H_b$ vanish.
First, in the case $g_1=g_2=g/2$, it is easy to see that the $b$-spin is effectively decoupled from the field-mode.
It means that, although the two spins are coupled with the field mode, within the $b$-space they evolve as the field mode were absent.
This condition simplifies $H_b$ as follows
\begin{equation}
    H_b={\varepsilon_b} \hat{\sigma}_{b}^{z} + {\gamma_{b}} \hat{\sigma}_{b}^{x} + \omega ~ \hat{a}^\dagger \hat{a},
\end{equation}
which, in turn, leads to a trivial spectrum consisting in the following infinite set of doublets $E_n^b=n \omega \pm \sqrt{\varepsilon_b^2 + \gamma_b^2}$.
$H_a$ is instead characterized by the `standard' single-spin QRM spectrum.
The first four eigenstates of the two-spin-mode system (given by the first two eigenvalues of $H_a$, $E_{0/1}^a$, and the first two eigenvalues of $H_b$, $E_{0/1}^b$) are shown in Fig. \ref{fig: spectrum}(a) for $\varepsilon_1=\varepsilon_2=0$ and $\gamma_x=2\gamma_y=0.4\omega$.

Another particular case corresponds to $\gamma_x=\gamma_y$; such a condition induces the emergence of a further constant of motion: $\hat{\Sigma}^z \equiv \hat{\sigma}_1^z + \hat{\sigma}_2^z$.
In this instance,
\begin{equation} \label{DQHO Ham}
    {H}_{a} =
{\varepsilon_a} \hat{\sigma}_{a}^{z} + \omega ~ \hat{a}^\dagger \hat{a} + g_a \left( \hat{a}^\dagger + \hat{a} \right)  \hat{\sigma}_{a}^z,
\end{equation}
can be reduced to the displaced quantum oscillator (DQHO) Hamiltonian since $\hat{\sigma}_a^z$ is a constant of motion.
The portion of spectrum of the two-spin-mode system stemming from $H_a$ consists in the infinite set of eigenvalues $(E_n^a)^\pm = \omega (n-\alpha^2) \pm \varepsilon_a$,
for $\sigma_a^z=\pm 1$ and with $\alpha=g_a/\omega$; the related eigenvectors turn out to be
\begin{equation}
\ket{\pm\pm}\otimes\ket{\mp\alpha,n}=\ket{\pm\pm}\otimes D(\mp\alpha)\ket{0},
\end{equation}
$D(\alpha)$ and $\ket{0}$ being the displacement operator and the vacuum state of the quantized bosonic mode, respectively \cite{Glauber69}.
In this case, the four lowest eigenvalues of the system seen before for $g_1=g_2=g/2$ ($E_{0/1}^a$, and $E_{0/1}^b$), when $\varepsilon_1=\varepsilon_2=0.25\omega$ and $\gamma_x=\gamma_y=0.3\omega$, are modified as shown in Fig. \ref{fig: spectrum}(b).

\textit{Unbiased QPT.}
Depending on the parameter-space region, the ground state (GS) of the two-spin-mode system \eqref{Hamiltonian} belongs to either the $a$ or $b$ space.
It can be derived by the GS of the effective quantum Rabi Hamiltonians which governs the dynamics in the $a$ ($H_a$) and $b$ ($H_b$) spaces, on the basis of the mapping in Eq. \eqref{Mapping}.

Let us first consider the unbiased case ($\varepsilon_1=\varepsilon_2=0$), assuming in addition equal couplings of the two spins with the mode, namely $g_1=g_2=g/2$.
This condition implies that the fictitious spin-1/2 $b$ is decoupled from the field mode.
In this instance, the two effective two-level Hamiltonians read indeed
\begin{equation}
\begin{aligned}
{H}_{a} = &
{\gamma_{a}} \hat{\sigma}_{a}^{x} +
\omega ~ \hat{a}^\dagger \hat{a} +
g \left( \hat{a}^\dagger + \hat{a} \right)  \hat{\sigma}_{a}^z, \\
{H}_{b} = &
{\gamma_{b}} \hat{\sigma}_{b}^{x} +
\omega ~ \hat{a}^\dagger \hat{a}.
\end{aligned}
\end{equation}
The ground energy of $H_b$ is trivial and corresponds to $E_0^b=-\gamma_b$, while $E_0^a$ can be derived analytically \cite{Braak11} (see the supplemental material).

\begin{figure}[] 
\begin{center}
{\includegraphics[width=0.22\textwidth]{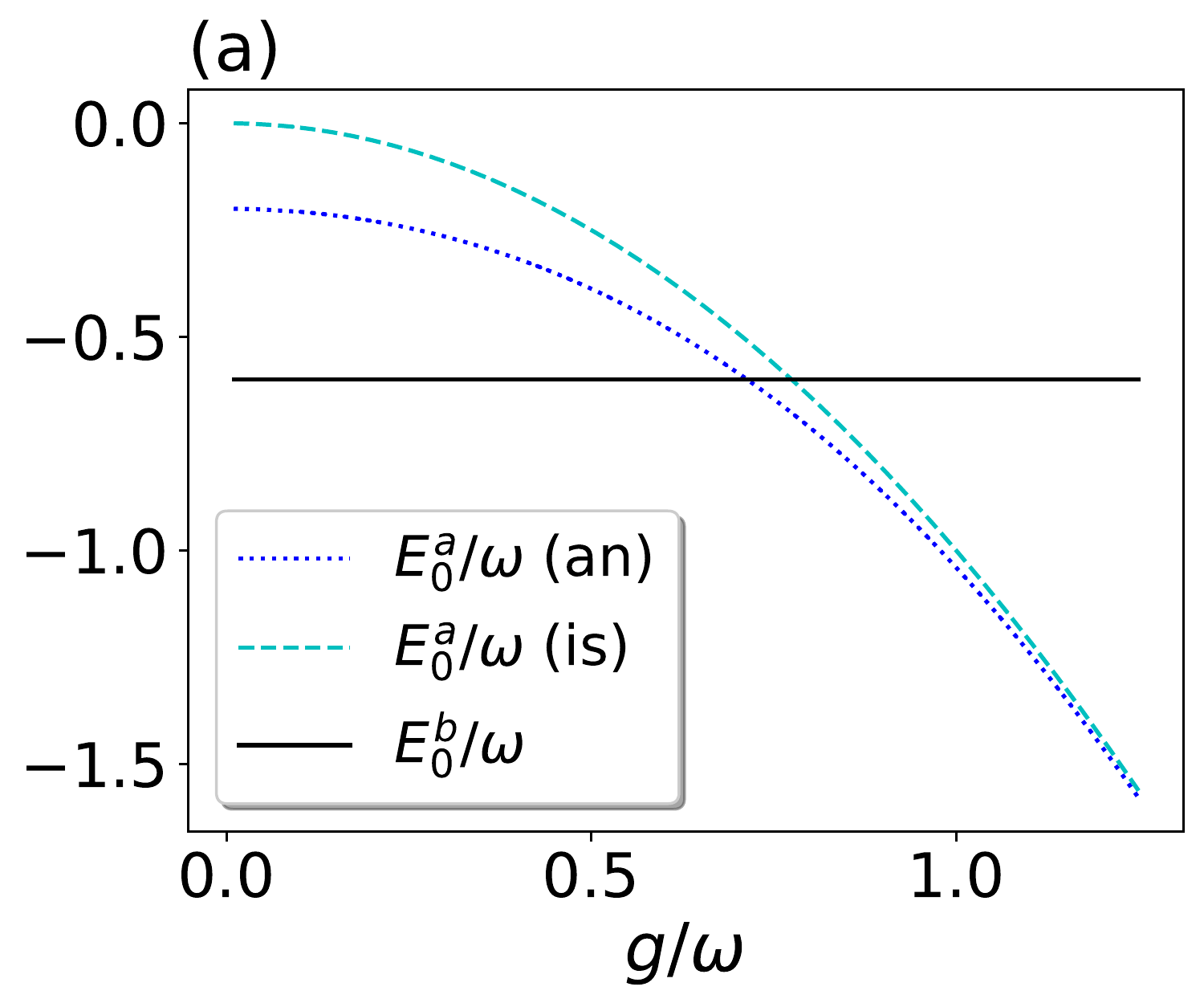}}
\captionsetup{justification=raggedright,format=plain,skip=4pt}%
{\includegraphics[width=0.22\textwidth]{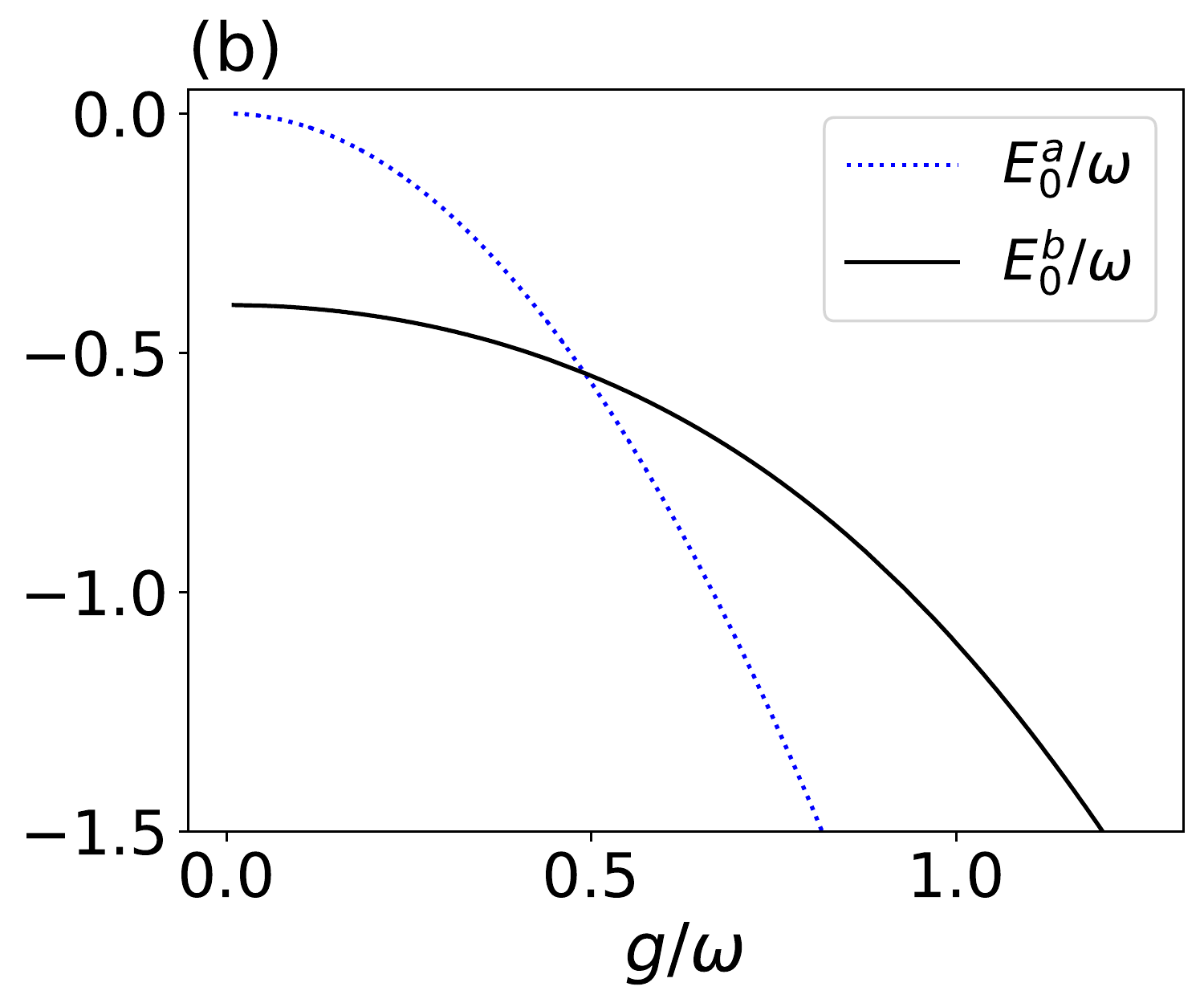}}
\caption{The lowest eigenvalue $E_0^b$ of $H_b$ (black solid line) and $E_0^a$ of $H_a$ for: (a) $\varepsilon_1=\varepsilon_2=0$, $g_1=g_2=g/2$ and in case of spin-spin coupling anisotropy ($\gamma_x=2\gamma_y=0.4\omega$, blue dotted line), and spin-spin coupling isotropy ($\gamma_x=\gamma_y=0.3\omega$, cyan dashed line);
(b) $\varepsilon_1=\varepsilon_2=0$, $2g_a/3=g_b=g$ and $\gamma_x=\gamma_y=0.2\omega$.}
\label{fig: qpt}
\end{center}
\end{figure}

By considering the anisotropic case $\gamma_x=2\gamma_y=0.4\omega$, a QPT happens at the critical value $g_c \approx 0.714$, as it is clearly shown in Fig. \ref{fig: qpt}(a) (solid black and blue dotted lines).
It means that the ground state of the two-spin-mode system is placed in the $b$-space for $g<g_c$ and corresponds to $\ket{GS_b} = (\ket{+-}-\ket{-+})/\sqrt{2} \otimes \ket{0}$.
As far as $g>g_c$, it `moves' into the $a$-space and can be written as \cite{Zhong13, Ying15}
\begin{equation}
    \ket{GS_a} = \ket{++} \otimes f_{1a}^-(\hat{a}^\dagger) \ket{0} - \ket{--} \otimes f_{2a}^-(\hat{a}^\dagger) \ket{0},
\end{equation}
where $f_{1a}^-$ and $f_{2a}^-$ are functions of the operator $\hat{a}^\dagger$ expressible in terms of the confluent Heun functions \footnote{see Eqs. 32, 33 and 34 in Ref. \cite{Zhong13}; our eigenstate looks different from the one in Ref. \cite{Zhong13} since our effective Hamiltonian $H_a$ (and $H_b$) is a QRM Hamiltonian rotated of $\pi/2$ around the $\hat{y}$-axis}. 
This quantum phase transition can be experimentally detected by measuring the net magnetization ($M_z=\average{GS_k|(\hat{\sigma}_1^z+\hat{\sigma}_2^z)/2|GS_k}$, $k=a,b$) of the two-spin system.
$\ket{GS_b}$, and more in general the whole $b$-space, are in fact characterized by a vanishing net spin magnetization since the two involved spin states are $\{ \ket{+-}, \ket{-+} \}$.
$\ket{GS_a}$, instead, presents, in general, a non-vanishing net spin magnetization.
This implies that, moving from the region $g<g_c$ to $g>g_c$, $M_z$ abruptly changes from a vanishing value to a non-vanishing one.
The highlighted phase transition can be then classified as a first-order quantum phase transition with the net spin magnetization as the order parameter.

It is worth noticing that in the isotropic case ($\gamma_x=\gamma_y=\gamma/2$), since $\gamma_a=0$, ${H}_{a}$ reduces to the DQHO Hamiltonian in Eq. \eqref{DQHO Ham} (with $\varepsilon_a=0$ for the unbiased case under scrutiny).
In this instance, the lowest-energy state of $H_a$ is doubly degenerate.
In terms of the two spins, by considering Eq. \eqref{Mapping}, they read:
\begin{equation} \label{GS DQHO}
  \ket{GS_a} = \left\{
    \begin{aligned}
      &\ket{++}\otimes D(-\alpha)\ket{0}, \\
      &\ket{--} \otimes D(\alpha)\ket{0}.
    \end{aligned}
  \right.
\end{equation}
Such a degeneracy is due to the existence of a further constant of motion: $\hat{\Sigma}^z$.
The related eigenvalue simply reads $E_0^a = -g^2/\omega$, while $E_0^b=-\gamma$ and the related lowest-energy state remain unchanged: $\ket{GS_b} = (\ket{+-}-\ket{-+})/\sqrt{2} \otimes \ket{0}$ (already mapped into the two-spin system).
In Fig. \ref{fig: qpt}(a) (cyan dashed line) it is possible to see that the quantum phase transition occurs as well, but the crossing point corresponds to a higher critical value of the coupling parameter, namely $g_c \approx 0.775$.
This shows that the isotropy level of the spin-spin coupling (which could depend on the geometry of the actual spin system) confers a different symmetry to the Hamiltonian.
This implies, in turn, the emergence of different physical features like, for example, the critical point which separates two different experimentally measurable phases.
We underline that, besides the magnetization, also the level of entanglement between the two qubits undergoes an abrupt change.
In this case, it is indeed easy to calculate the concurrence $C$ \cite{Wootters98}, which vanishes for $\ket{GS_a}$ and is maximum ($C=1$) for $\ket{GS_b}$. 

In Fig. \ref{fig: qpt}(b) we show the QPT both for the isotropic case $\gamma_x=\gamma_y=\gamma/2=0.2\omega$ and for a different coupling to the mode of the two spins ($2g_a/3=g_b=g$).
In this scenario the two Hamiltonians read
\begin{equation}
\begin{aligned}
{H}_{a} = &
\omega ~ \hat{a}^\dagger \hat{a} +
g_a \left( \hat{a}^\dagger + \hat{a} \right)  \hat{\sigma}_{a}^z, \\
{H}_{b} = &
\gamma \hat{\sigma}_{b}^{x} +
\omega ~ \hat{a}^\dagger \hat{a}+
g_b \left( \hat{a}^\dagger + \hat{a} \right)  \hat{\sigma}_{b}^z.
\end{aligned}
\end{equation}
In this case $E_0^b$ coincides with the energy of the lowest-energy state of the related QRM, which in terms of the two spin reads
\begin{equation}
    \ket{GS_b} = \ket{+-} \otimes f_{1b}^-(\hat{a}^\dagger) \ket{0} - \ket{-+} \otimes f_{2b}^-(\hat{a}^\dagger) \ket{0},
\end{equation}
while $E_0^a$ is the energy of the lowest-energy state of the DQHO given in Eq. \eqref{GS DQHO}.
The QPT at $g_c \approx 0.491$ is due to the different dependence of $E_0^a$ and $E_0^b$ on $g$.

We remark that the previously highlighted QPTs do not depend on any external parameter (e.g. the applied magnetic field).
Rather, they only depend on the relative weights of the internal parameters characterizing the system: the spin-spin coupling, the spin-mode coupling and the mode energy.
Therefore, such a phenomenon can be interpreted as a self organization of the system: depending on the geometry as well as on the physical features of the system, the latter can exhibit different phases for the GS.
As seen before, in fact, different couplings and related anisotropies deeply determine the critical value for which a QPT occurs and consequently different structures of the phase space.
In nuclear magnetic resonance, for example, the typical range of the spin-spin coupling is 10-300 Hz, depending on the molecule \cite{Vandersypen05}.
The interaction strength can instead reach the kHz range in microwave-driven trapped ion scenarios \cite{Weidt16}.
Further, thanks to the huge electric-dipole moments of the Rydberg states, the effective spin-spin coupling in Rydberg atoms and ions can reach a few MHz \cite{Gaetan09, Urban09}.

\begin{figure}[t!!] 
\begin{center}
{\includegraphics[width=0.22\textwidth]{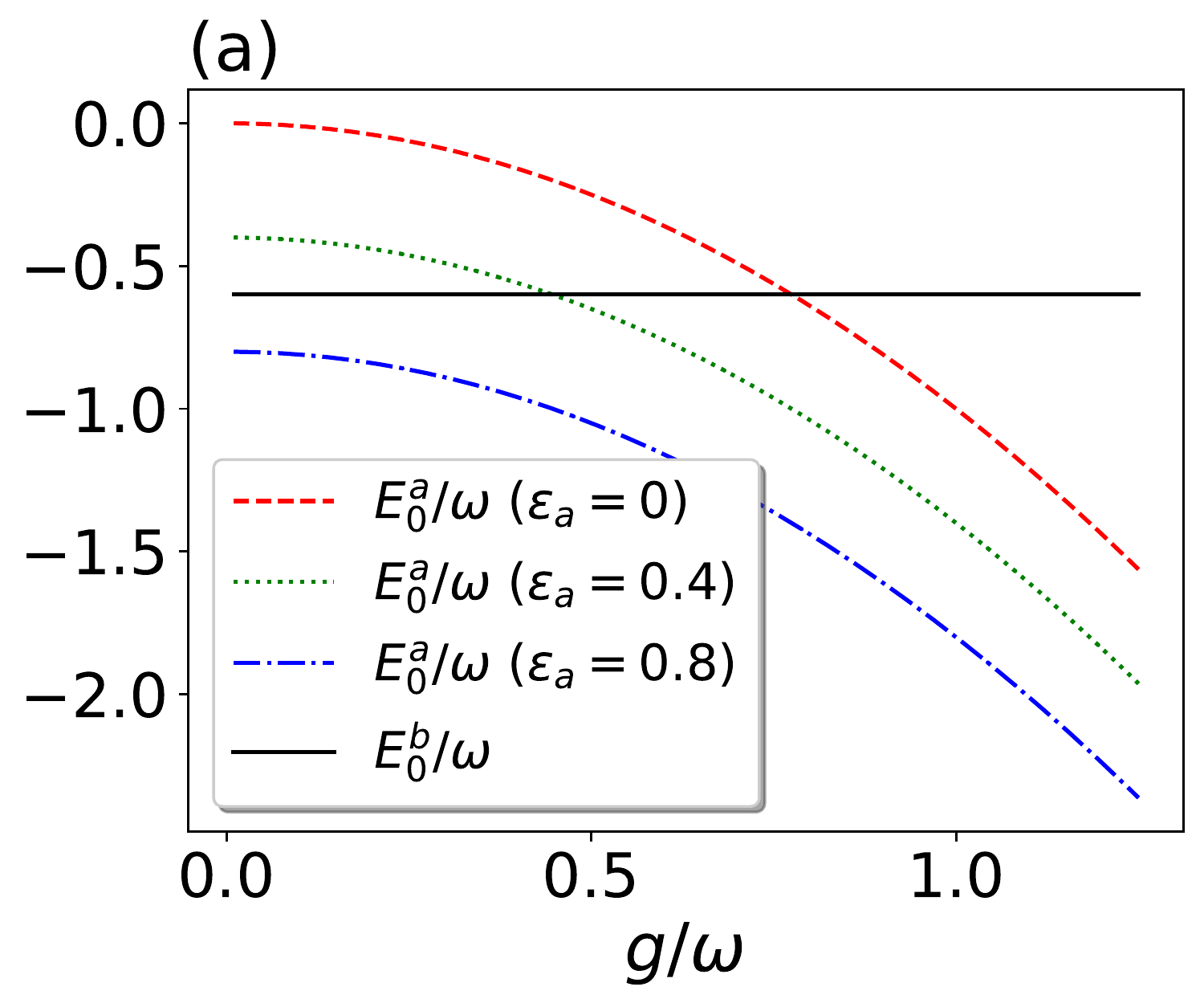}}
{\includegraphics[width=0.22\textwidth]{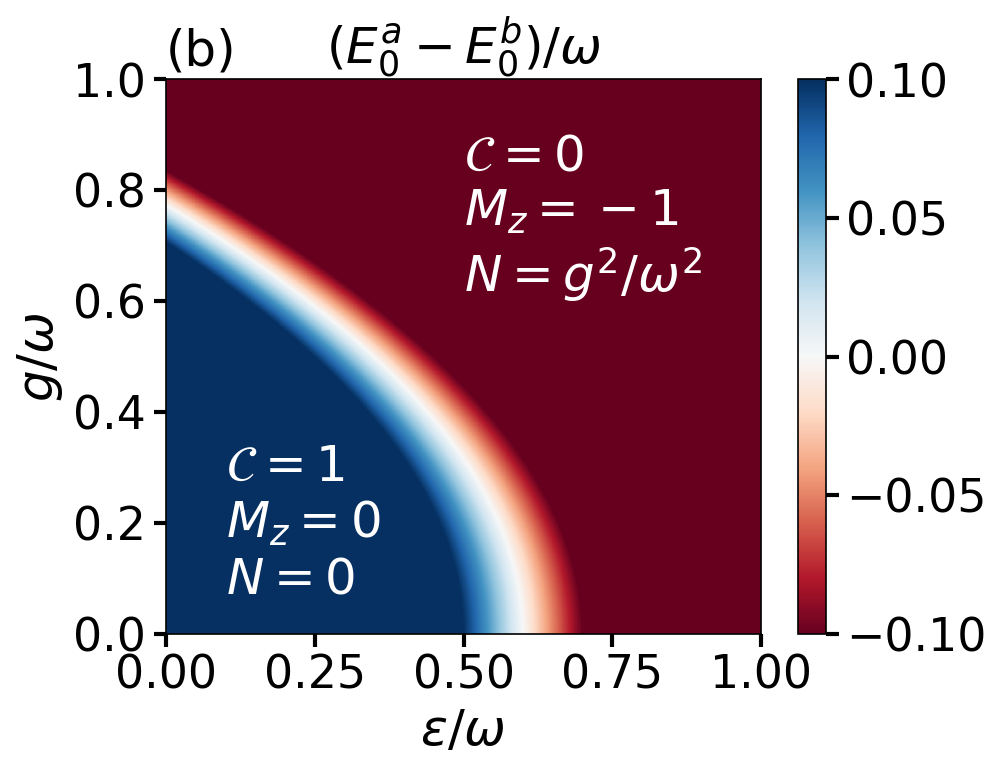}}
\captionsetup{justification=raggedright,format=plain,skip=4pt}
\caption{(a) The lowest eigenvalue $E_0^b$ of $H_b$ (black solid line) and $E_0^a$ of $H_a$ for $g_1=g_2=g/2$, $\gamma_x=\gamma_y=0.3\omega$, and $\varepsilon_1=\varepsilon_2=0$ (red dashed line), $\varepsilon_1=\varepsilon_2=0.2\omega$ (green dotted line), $\varepsilon_1=\varepsilon_2=0.4\omega$ (blue dot-dashed line).
(b) The normalized energy difference $(E_0^a-E_0^b)/\omega$ in the $g$-$\varepsilon$ space. The two phases are characterized by different values of the two-spin magnetization $M_z$, the mean photon number $N$, and the concurrence $C$.}
\label{fig: qpt iso mf}
\end{center}
\end{figure}

\textit{Biased QPT.}
In order to study the effects stemming from the presence of bias terms, let us consider a homogeneous ($\varepsilon_1=\varepsilon_2=\varepsilon/2$) magnetic field applied to the two spins and homogeneous spin-mode coupling ($g_1=g_2=g/2$) and spin-spin-coupling ($\gamma_x=\gamma_y=\gamma/2$). 
In this instance the two effective Hamiltonians read
\begin{subequations} \label{Ha Hb Biased QPT}
\begin{align}
{H}_{a} = &
\varepsilon \hat{\sigma}_{a}^{z} +
\omega ~ \hat{a}^\dagger \hat{a} +
g \left( \hat{a}^\dagger + \hat{a} \right)  \hat{\sigma}_{a}^z, \\
{H}_{b} = &
\gamma \hat{\sigma}_{b}^{x} +
\omega ~ \hat{a}^\dagger \hat{a}.
\end{align}
\end{subequations}
The magnetic field removes the degeneracy of the lowest-energy state of $H_a$, which this time [mapped through \eqref{Mapping}] results to be $\ket{GS_a}=\ket{--} \otimes D(\alpha)\ket{0}$ with $E_0^a=-g^2/\omega - \varepsilon$ as ground energy.
The (mapped) lowest-energy state of $H_b$, instead, is again $\ket{GS_b} = (\ket{+-}-\ket{-+})/\sqrt{2} \otimes \ket{0}$ with eigenvalue $E_0^b=-\gamma$.

In Fig. \ref{fig: qpt iso mf}(a) $E_0^a$ is plotted versus $g/\omega$ for different values of $\varepsilon$.
We see that, depending on the value of the homogeneous magnetic field, the critical value of $g$ corresponding to a QPT is different (red dashed and green dotted lines).
In particular, when $\varepsilon > \gamma$ the critical point does not exist, implying that no QPT occurs (blue dot-dashed line).
This aspect suggests that, fixing the value of the spin-mode coupling, an $\varepsilon$-dependent QPT is present.
In Fig. \ref{fig: qpt iso mf}(b), indeed, by plotting the difference $E_0^a-E_0^b$ in the $(\varepsilon,g)$ space, two distinct regions corresponding to different phases of the system clearly emerge.
The white strip coincides with the critical points where the QPT occurs.

It is worth noticing that these two phases are characterized by a different level of entanglement between the two qubits.
In $\ket{GS_a}$ the latter are indeed in a disentangled state, whereas $\ket{GS_b}$ exhibits a maximally entangled state of the two qubits.
It means that the QPT is characterized by an abrupt change of the concurrence ${C}$ \cite{Wootters98}, namely ${C}=0$ and ${C}=1$ in the subspace $a$ and $b$, respectively [see Fig. \ref{fig: qpt iso mf}(b)].
The physical reason at the basis of such an effect is that the spin-spin coupling, responsible for the two-qubit entanglement, is present in the effective Hamiltonian $H_b$, while it is absent in $H_a$ [see Eqs. \eqref{Ha Hb Biased QPT}].
Moreover, also the two-qubit magnetization $M_z$ and the mean photon number $N = \average{GS_k|\hat{a}^\dagger a|GS_k}$, $(k=a,b)$ exhibit a discontinuous behaviour in the QPT: $\{M_z=-1, ~ N=\alpha^2=g^2/\omega^2 \}$ and $\{M_z=0, ~ N=0 \}$ for $\ket{GS_a}$ and $\ket{GS_b}$, respectively.
Therefore, the concurrence, the spin magnetization and the mean photon number results to be order parameters of this first-order QPT.

Finally, we underline that an analogous, but qualitatively different, QPT occurs by fixing the spin-mode coupling $g$ and varying the spin-spin coupling $\gamma$.
This shows that the two-qubit QRM here analysed admits QPTs not as a particular case, that is for specific conditions on the Hamiltonian parameters, rather a wide range of different scenarios exists where QPTs are allowed.

\textit{Conclusions.}
The two-qubit QRM here investigated allows to focus on the effects stemming from the interplay between the qubit-qubit coupling and the qubit-mode coupling.
The two-qubit-mode problem can be exactly reduced into two independent single-qubit-mode (sub)problems, which turns out to be integrable.
This circumstance allows to analytically find the occurrence of QPTs originating discontinuities in the spin magnetization, the mean photon number and the concurrence.
Our exact approach \cite{GMIV, GMV2, GMMM, GLSM, GBNM, GVdCVM, GdCMV} can be applied to other scenarios paving the way for further investigations and applications.

\bibliography{biblio_qrm.bib}

\begin{thebibliography}{68}%
\makeatletter
\providecommand \@ifxundefined [1]{%
 \@ifx{#1\undefined}
}%
\providecommand \@ifnum [1]{%
 \ifnum #1\expandafter \@firstoftwo
 \else \expandafter \@secondoftwo
 \fi
}%
\providecommand \@ifx [1]{%
 \ifx #1\expandafter \@firstoftwo
 \else \expandafter \@secondoftwo
 \fi
}%
\providecommand \natexlab [1]{#1}%
\providecommand \enquote  [1]{``#1''}%
\providecommand \bibnamefont  [1]{#1}%
\providecommand \bibfnamefont [1]{#1}%
\providecommand \citenamefont [1]{#1}%
\providecommand \href@noop [0]{\@secondoftwo}%
\providecommand \href [0]{\begingroup \@sanitize@url \@href}%
\providecommand \@href[1]{\@@startlink{#1}\@@href}%
\providecommand \@@href[1]{\endgroup#1\@@endlink}%
\providecommand \@sanitize@url [0]{\catcode `\\12\catcode `\$12\catcode
  `\&12\catcode `\#12\catcode `\^12\catcode `\_12\catcode `\%12\relax}%
\providecommand \@@startlink[1]{}%
\providecommand \@@endlink[0]{}%
\providecommand \url  [0]{\begingroup\@sanitize@url \@url }%
\providecommand \@url [1]{\endgroup\@href {#1}{\urlprefix }}%
\providecommand \urlprefix  [0]{URL }%
\providecommand \Eprint [0]{\href }%
\providecommand \doibase [0]{http://dx.doi.org/}%
\providecommand \selectlanguage [0]{\@gobble}%
\providecommand \bibinfo  [0]{\@secondoftwo}%
\providecommand \bibfield  [0]{\@secondoftwo}%
\providecommand \translation [1]{[#1]}%
\providecommand \BibitemOpen [0]{}%
\providecommand \bibitemStop [0]{}%
\providecommand \bibitemNoStop [0]{.\EOS\space}%
\providecommand \EOS [0]{\spacefactor3000\relax}%
\providecommand \BibitemShut  [1]{\csname bibitem#1\endcsname}%
\let\auto@bib@innerbib\@empty
\bibitem [{\citenamefont {Rabi}(1936)}]{Rabi36}%
  \BibitemOpen
  \bibfield  {author} {\bibinfo {author} {\bibfnamefont {I.~I.}\ \bibnamefont
  {Rabi}},\ }\href {\doibase 10.1103/PhysRev.49.324} {\bibfield  {journal}
  {\bibinfo  {journal} {Phys. Rev.}\ }\textbf {\bibinfo {volume} {49}},\
  \bibinfo {pages} {324} (\bibinfo {year} {1936})}\BibitemShut {NoStop}%
\bibitem [{\citenamefont {Rabi}(1937)}]{Rabi37}%
  \BibitemOpen
  \bibfield  {author} {\bibinfo {author} {\bibfnamefont {I.~I.}\ \bibnamefont
  {Rabi}},\ }\href {\doibase 10.1103/PhysRev.51.652} {\bibfield  {journal}
  {\bibinfo  {journal} {Phys. Rev.}\ }\textbf {\bibinfo {volume} {51}},\
  \bibinfo {pages} {652} (\bibinfo {year} {1937})}\BibitemShut {NoStop}%
\bibitem [{\citenamefont {Jaynes}\ and\ \citenamefont {Cummings}(1963)}]{JC}%
  \BibitemOpen
  \bibfield  {author} {\bibinfo {author} {\bibfnamefont {E.}~\bibnamefont
  {Jaynes}}\ and\ \bibinfo {author} {\bibfnamefont {F.}~\bibnamefont
  {Cummings}},\ }\href {\doibase 10.1109/PROC.1963.1664} {\bibfield  {journal}
  {\bibinfo  {journal} {Proceedings of the IEEE}\ }\textbf {\bibinfo {volume}
  {51}},\ \bibinfo {pages} {89} (\bibinfo {year} {1963})}\BibitemShut {NoStop}%
\bibitem [{\citenamefont {Braak}\ \emph {et~al.}(2016)\citenamefont {Braak},
  \citenamefont {Chen}, \citenamefont {Batchelor},\ and\ \citenamefont
  {Solano}}]{Braak16}%
  \BibitemOpen
  \bibfield  {author} {\bibinfo {author} {\bibfnamefont {D.}~\bibnamefont
  {Braak}}, \bibinfo {author} {\bibfnamefont {Q.-H.}\ \bibnamefont {Chen}},
  \bibinfo {author} {\bibfnamefont {M.~T.}\ \bibnamefont {Batchelor}}, \ and\
  \bibinfo {author} {\bibfnamefont {E.}~\bibnamefont {Solano}},\ }\href
  {\doibase 10.1088/1751-8113/49/30/300301} {\bibfield  {journal} {\bibinfo
  {journal} {Journal of Physics A: Mathematical and Theoretical}\ }\textbf
  {\bibinfo {volume} {49}},\ \bibinfo {pages} {300301} (\bibinfo {year}
  {2016})}\BibitemShut {NoStop}%
\bibitem [{\citenamefont {Braak}(2011)}]{Braak11}%
  \BibitemOpen
  \bibfield  {author} {\bibinfo {author} {\bibfnamefont {D.}~\bibnamefont
  {Braak}},\ }\href {\doibase 10.1103/PhysRevLett.107.100401} {\bibfield
  {journal} {\bibinfo  {journal} {Phys. Rev. Lett.}\ }\textbf {\bibinfo
  {volume} {107}},\ \bibinfo {pages} {100401} (\bibinfo {year}
  {2011})}\BibitemShut {NoStop}%
\bibitem [{\citenamefont {Ashhab}(2013)}]{Ashhab13}%
  \BibitemOpen
  \bibfield  {author} {\bibinfo {author} {\bibfnamefont {S.}~\bibnamefont
  {Ashhab}},\ }\href {\doibase 10.1103/PhysRevA.87.013826} {\bibfield
  {journal} {\bibinfo  {journal} {Phys. Rev. A}\ }\textbf {\bibinfo {volume}
  {87}},\ \bibinfo {pages} {013826} (\bibinfo {year} {2013})}\BibitemShut
  {NoStop}%
\bibitem [{\citenamefont {Hwang}\ \emph {et~al.}(2015)\citenamefont {Hwang},
  \citenamefont {Puebla},\ and\ \citenamefont {Plenio}}]{Wang15}%
  \BibitemOpen
  \bibfield  {author} {\bibinfo {author} {\bibfnamefont {M.-J.}\ \bibnamefont
  {Hwang}}, \bibinfo {author} {\bibfnamefont {R.}~\bibnamefont {Puebla}}, \
  and\ \bibinfo {author} {\bibfnamefont {M.~B.}\ \bibnamefont {Plenio}},\
  }\href {\doibase 10.1103/PhysRevLett.115.180404} {\bibfield  {journal}
  {\bibinfo  {journal} {Phys. Rev. Lett.}\ }\textbf {\bibinfo {volume} {115}},\
  \bibinfo {pages} {180404} (\bibinfo {year} {2015})}\BibitemShut {NoStop}%
\bibitem [{\citenamefont {Ying}\ \emph {et~al.}(2015)\citenamefont {Ying},
  \citenamefont {Liu}, \citenamefont {Luo}, \citenamefont {Lin},\ and\
  \citenamefont {You}}]{Ying15}%
  \BibitemOpen
  \bibfield  {author} {\bibinfo {author} {\bibfnamefont {Z.-J.}\ \bibnamefont
  {Ying}}, \bibinfo {author} {\bibfnamefont {M.}~\bibnamefont {Liu}}, \bibinfo
  {author} {\bibfnamefont {H.-G.}\ \bibnamefont {Luo}}, \bibinfo {author}
  {\bibfnamefont {H.-Q.}\ \bibnamefont {Lin}}, \ and\ \bibinfo {author}
  {\bibfnamefont {J.~Q.}\ \bibnamefont {You}},\ }\href {\doibase
  10.1103/PhysRevA.92.053823} {\bibfield  {journal} {\bibinfo  {journal} {Phys.
  Rev. A}\ }\textbf {\bibinfo {volume} {92}},\ \bibinfo {pages} {053823}
  (\bibinfo {year} {2015})}\BibitemShut {NoStop}%
\bibitem [{\citenamefont {Ying}(2022{\natexlab{a}})}]{Ying22}%
  \BibitemOpen
  \bibfield  {author} {\bibinfo {author} {\bibfnamefont {Z.-J.}\ \bibnamefont
  {Ying}},\ }\href {\doibase https://doi.org/10.1002/qute.202100088} {\bibfield
   {journal} {\bibinfo  {journal} {Advanced Quantum Technologies}\ }\textbf
  {\bibinfo {volume} {5}},\ \bibinfo {pages} {2100088} (\bibinfo {year}
  {2022}{\natexlab{a}})}\BibitemShut {NoStop}%
\bibitem [{\citenamefont {Ying}(2022{\natexlab{b}})}]{Ying22new}%
  \BibitemOpen
  \bibfield  {author} {\bibinfo {author} {\bibfnamefont {Z.-J.}\ \bibnamefont
  {Ying}},\ }\href {\doibase https://doi.org/10.1002/qute.202100165} {\bibfield
   {journal} {\bibinfo  {journal} {Advanced Quantum Technologies}\ }\textbf
  {\bibinfo {volume} {5}},\ \bibinfo {pages} {2100165} (\bibinfo {year}
  {2022}{\natexlab{b}})}\BibitemShut {NoStop}%
\bibitem [{\citenamefont {Vojta}\ \emph {et~al.}(2005)\citenamefont {Vojta},
  \citenamefont {Tong},\ and\ \citenamefont {Bulla}}]{Vojta05}%
  \BibitemOpen
  \bibfield  {author} {\bibinfo {author} {\bibfnamefont {M.}~\bibnamefont
  {Vojta}}, \bibinfo {author} {\bibfnamefont {N.-H.}\ \bibnamefont {Tong}}, \
  and\ \bibinfo {author} {\bibfnamefont {R.}~\bibnamefont {Bulla}},\ }\href
  {\doibase 10.1103/PhysRevLett.94.070604} {\bibfield  {journal} {\bibinfo
  {journal} {Phys. Rev. Lett.}\ }\textbf {\bibinfo {volume} {94}},\ \bibinfo
  {pages} {070604} (\bibinfo {year} {2005})}\BibitemShut {NoStop}%
\bibitem [{\citenamefont {Carollo}\ \emph {et~al.}(2020)\citenamefont
  {Carollo}, \citenamefont {Valenti},\ and\ \citenamefont
  {Spagnolo}}]{Carollo20}%
  \BibitemOpen
  \bibfield  {author} {\bibinfo {author} {\bibfnamefont {A.}~\bibnamefont
  {Carollo}}, \bibinfo {author} {\bibfnamefont {D.}~\bibnamefont {Valenti}}, \
  and\ \bibinfo {author} {\bibfnamefont {B.}~\bibnamefont {Spagnolo}},\ }\href
  {\doibase https://doi.org/10.1016/j.physrep.2019.11.002} {\bibfield
  {journal} {\bibinfo  {journal} {Phys. Rep.}\ }\textbf {\bibinfo {volume}
  {838}},\ \bibinfo {pages} {1} (\bibinfo {year} {2020})}\BibitemShut {NoStop}%
\bibitem [{\citenamefont {Rossini}\ and\ \citenamefont
  {Vicari}(2021)}]{Rossini21}%
  \BibitemOpen
  \bibfield  {author} {\bibinfo {author} {\bibfnamefont {D.}~\bibnamefont
  {Rossini}}\ and\ \bibinfo {author} {\bibfnamefont {E.}~\bibnamefont
  {Vicari}},\ }\href {\doibase https://doi.org/10.1016/j.physrep.2021.08.003}
  {\bibfield  {journal} {\bibinfo  {journal} {Phys. Rep.}\ }\textbf {\bibinfo
  {volume} {936}},\ \bibinfo {pages} {1} (\bibinfo {year} {2021})},\ \bibinfo
  {note} {coherent and dissipative dynamics at quantum phase
  transitions}\BibitemShut {NoStop}%
\bibitem [{\citenamefont {Barenco}\ \emph {et~al.}(1995)\citenamefont
  {Barenco}, \citenamefont {Bennett}, \citenamefont {Cleve}, \citenamefont
  {DiVincenzo}, \citenamefont {Margolus}, \citenamefont {Shor}, \citenamefont
  {Sleator}, \citenamefont {Smolin},\ and\ \citenamefont
  {Weinfurter}}]{Barenco95}%
  \BibitemOpen
  \bibfield  {author} {\bibinfo {author} {\bibfnamefont {A.}~\bibnamefont
  {Barenco}}, \bibinfo {author} {\bibfnamefont {C.~H.}\ \bibnamefont
  {Bennett}}, \bibinfo {author} {\bibfnamefont {R.}~\bibnamefont {Cleve}},
  \bibinfo {author} {\bibfnamefont {D.~P.}\ \bibnamefont {DiVincenzo}},
  \bibinfo {author} {\bibfnamefont {N.}~\bibnamefont {Margolus}}, \bibinfo
  {author} {\bibfnamefont {P.}~\bibnamefont {Shor}}, \bibinfo {author}
  {\bibfnamefont {T.}~\bibnamefont {Sleator}}, \bibinfo {author} {\bibfnamefont
  {J.~A.}\ \bibnamefont {Smolin}}, \ and\ \bibinfo {author} {\bibfnamefont
  {H.}~\bibnamefont {Weinfurter}},\ }\href {\doibase 10.1103/PhysRevA.52.3457}
  {\bibfield  {journal} {\bibinfo  {journal} {Phys. Rev. A}\ }\textbf {\bibinfo
  {volume} {52}},\ \bibinfo {pages} {3457} (\bibinfo {year}
  {1995})}\BibitemShut {NoStop}%
\bibitem [{\citenamefont {Hua}\ \emph {et~al.}(2014)\citenamefont {Hua},
  \citenamefont {Tao},\ and\ \citenamefont {Deng}}]{Hua14}%
  \BibitemOpen
  \bibfield  {author} {\bibinfo {author} {\bibfnamefont {M.}~\bibnamefont
  {Hua}}, \bibinfo {author} {\bibfnamefont {M.-J.}\ \bibnamefont {Tao}}, \ and\
  \bibinfo {author} {\bibfnamefont {F.-G.}\ \bibnamefont {Deng}},\ }\href
  {\doibase 10.1103/PhysRevA.90.012328} {\bibfield  {journal} {\bibinfo
  {journal} {Phys. Rev. A}\ }\textbf {\bibinfo {volume} {90}},\ \bibinfo
  {pages} {012328} (\bibinfo {year} {2014})}\BibitemShut {NoStop}%
\bibitem [{\citenamefont {Romero}\ \emph {et~al.}(2012)\citenamefont {Romero},
  \citenamefont {Ballester}, \citenamefont {Wang}, \citenamefont {Scarani},\
  and\ \citenamefont {Solano}}]{Romero12}%
  \BibitemOpen
  \bibfield  {author} {\bibinfo {author} {\bibfnamefont {G.}~\bibnamefont
  {Romero}}, \bibinfo {author} {\bibfnamefont {D.}~\bibnamefont {Ballester}},
  \bibinfo {author} {\bibfnamefont {Y.~M.}\ \bibnamefont {Wang}}, \bibinfo
  {author} {\bibfnamefont {V.}~\bibnamefont {Scarani}}, \ and\ \bibinfo
  {author} {\bibfnamefont {E.}~\bibnamefont {Solano}},\ }\href {\doibase
  10.1103/PhysRevLett.108.120501} {\bibfield  {journal} {\bibinfo  {journal}
  {Phys. Rev. Lett.}\ }\textbf {\bibinfo {volume} {108}},\ \bibinfo {pages}
  {120501} (\bibinfo {year} {2012})}\BibitemShut {NoStop}%
\bibitem [{\citenamefont {Barends}\ and\ \citenamefont {\textit{et
  al.}}(2019)}]{Barends19}%
  \BibitemOpen
  \bibfield  {author} {\bibinfo {author} {\bibfnamefont {R.}~\bibnamefont
  {Barends}}\ and\ \bibinfo {author} {\bibnamefont {\textit{et al.}}},\ }\href
  {\doibase 10.1103/PhysRevLett.123.210501} {\bibfield  {journal} {\bibinfo
  {journal} {Phys. Rev. Lett.}\ }\textbf {\bibinfo {volume} {123}},\ \bibinfo
  {pages} {210501} (\bibinfo {year} {2019})}\BibitemShut {NoStop}%
\bibitem [{\citenamefont {Kang}\ \emph {et~al.}(2016)\citenamefont {Kang},
  \citenamefont {Chen}, \citenamefont {Wu}, \citenamefont {Huang},
  \citenamefont {Song},\ and\ \citenamefont {Xia}}]{Kang16}%
  \BibitemOpen
  \bibfield  {author} {\bibinfo {author} {\bibfnamefont {Y.-H.}\ \bibnamefont
  {Kang}}, \bibinfo {author} {\bibfnamefont {Y.-H.}\ \bibnamefont {Chen}},
  \bibinfo {author} {\bibfnamefont {Q.-C.}\ \bibnamefont {Wu}}, \bibinfo
  {author} {\bibfnamefont {B.-H.}\ \bibnamefont {Huang}}, \bibinfo {author}
  {\bibfnamefont {J.}~\bibnamefont {Song}}, \ and\ \bibinfo {author}
  {\bibfnamefont {Y.}~\bibnamefont {Xia}},\ }\href@noop {} {\bibfield
  {journal} {\bibinfo  {journal} {Sci. Rep.}\ }\textbf {\bibinfo {volume}
  {6}},\ \bibinfo {pages} {1} (\bibinfo {year} {2016})}\BibitemShut {NoStop}%
\bibitem [{\citenamefont {Lu}\ \emph {et~al.}(2013)\citenamefont {Lu},
  \citenamefont {Xia}, \citenamefont {Song},\ and\ \citenamefont {An}}]{Lu13}%
  \BibitemOpen
  \bibfield  {author} {\bibinfo {author} {\bibfnamefont {M.}~\bibnamefont
  {Lu}}, \bibinfo {author} {\bibfnamefont {Y.}~\bibnamefont {Xia}}, \bibinfo
  {author} {\bibfnamefont {J.}~\bibnamefont {Song}}, \ and\ \bibinfo {author}
  {\bibfnamefont {N.~B.}\ \bibnamefont {An}},\ }\href {\doibase
  10.1364/JOSAB.30.002142} {\bibfield  {journal} {\bibinfo  {journal} {J. Opt.
  Soc. Am. B}\ }\textbf {\bibinfo {volume} {30}},\ \bibinfo {pages} {2142}
  (\bibinfo {year} {2013})}\BibitemShut {NoStop}%
\bibitem [{\citenamefont {Li}\ and\ \citenamefont {Paraoanu}(2009)}]{Li09}%
  \BibitemOpen
  \bibfield  {author} {\bibinfo {author} {\bibfnamefont {J.}~\bibnamefont
  {Li}}\ and\ \bibinfo {author} {\bibfnamefont {G.~S.}\ \bibnamefont
  {Paraoanu}},\ }\href {\doibase 10.1088/1367-2630/11/11/113020} {\bibfield
  {journal} {\bibinfo  {journal} {New Journal of Physics}\ }\textbf {\bibinfo
  {volume} {11}},\ \bibinfo {pages} {113020} (\bibinfo {year}
  {2009})}\BibitemShut {NoStop}%
\bibitem [{\citenamefont {Dicke}(1954)}]{Dicke54}%
  \BibitemOpen
  \bibfield  {author} {\bibinfo {author} {\bibfnamefont {R.~H.}\ \bibnamefont
  {Dicke}},\ }\href {\doibase 10.1103/PhysRev.93.99} {\bibfield  {journal}
  {\bibinfo  {journal} {Phys. Rev.}\ }\textbf {\bibinfo {volume} {93}},\
  \bibinfo {pages} {99} (\bibinfo {year} {1954})}\BibitemShut {NoStop}%
\bibitem [{\citenamefont {Chen}\ \emph {et~al.}(2012)\citenamefont {Chen},
  \citenamefont {Wang}, \citenamefont {He}, \citenamefont {Liu},\ and\
  \citenamefont {Wang}}]{Chen12}%
  \BibitemOpen
  \bibfield  {author} {\bibinfo {author} {\bibfnamefont {Q.-H.}\ \bibnamefont
  {Chen}}, \bibinfo {author} {\bibfnamefont {C.}~\bibnamefont {Wang}}, \bibinfo
  {author} {\bibfnamefont {S.}~\bibnamefont {He}}, \bibinfo {author}
  {\bibfnamefont {T.}~\bibnamefont {Liu}}, \ and\ \bibinfo {author}
  {\bibfnamefont {K.-L.}\ \bibnamefont {Wang}},\ }\href {\doibase
  10.1103/PhysRevA.86.023822} {\bibfield  {journal} {\bibinfo  {journal} {Phys.
  Rev. A}\ }\textbf {\bibinfo {volume} {86}},\ \bibinfo {pages} {023822}
  (\bibinfo {year} {2012})}\BibitemShut {NoStop}%
\bibitem [{\citenamefont {Felicetti}\ \emph {et~al.}(2015)\citenamefont
  {Felicetti}, \citenamefont {Pedernales}, \citenamefont {Egusquiza},
  \citenamefont {Romero}, \citenamefont {Lamata}, \citenamefont {Braak},\ and\
  \citenamefont {Solano}}]{Felicetti15}%
  \BibitemOpen
  \bibfield  {author} {\bibinfo {author} {\bibfnamefont {S.}~\bibnamefont
  {Felicetti}}, \bibinfo {author} {\bibfnamefont {J.~S.}\ \bibnamefont
  {Pedernales}}, \bibinfo {author} {\bibfnamefont {I.~L.}\ \bibnamefont
  {Egusquiza}}, \bibinfo {author} {\bibfnamefont {G.}~\bibnamefont {Romero}},
  \bibinfo {author} {\bibfnamefont {L.}~\bibnamefont {Lamata}}, \bibinfo
  {author} {\bibfnamefont {D.}~\bibnamefont {Braak}}, \ and\ \bibinfo {author}
  {\bibfnamefont {E.}~\bibnamefont {Solano}},\ }\href {\doibase
  10.1103/PhysRevA.92.033817} {\bibfield  {journal} {\bibinfo  {journal} {Phys.
  Rev. A}\ }\textbf {\bibinfo {volume} {92}},\ \bibinfo {pages} {033817}
  (\bibinfo {year} {2015})}\BibitemShut {NoStop}%
\bibitem [{\citenamefont {Zhang}(2013)}]{Zhang13}%
  \BibitemOpen
  \bibfield  {author} {\bibinfo {author} {\bibfnamefont {Y.-Z.}\ \bibnamefont
  {Zhang}},\ }\href {\doibase 10.1063/1.4826356} {\bibfield  {journal}
  {\bibinfo  {journal} {Journal of Mathematical Physics}\ }\textbf {\bibinfo
  {volume} {54}},\ \bibinfo {pages} {102104} (\bibinfo {year}
  {2013})}\BibitemShut {NoStop}%
\bibitem [{\citenamefont {Albert}(2012)}]{Albert12}%
  \BibitemOpen
  \bibfield  {author} {\bibinfo {author} {\bibfnamefont {V.~V.}\ \bibnamefont
  {Albert}},\ }\href {\doibase 10.1103/PhysRevLett.108.180401} {\bibfield
  {journal} {\bibinfo  {journal} {Phys. Rev. Lett.}\ }\textbf {\bibinfo
  {volume} {108}},\ \bibinfo {pages} {180401} (\bibinfo {year}
  {2012})}\BibitemShut {NoStop}%
\bibitem [{\citenamefont {Agarwal}\ \emph {et~al.}(2012)\citenamefont
  {Agarwal}, \citenamefont {Rafsanjani},\ and\ \citenamefont
  {Eberly}}]{Agarwal12}%
  \BibitemOpen
  \bibfield  {author} {\bibinfo {author} {\bibfnamefont {S.}~\bibnamefont
  {Agarwal}}, \bibinfo {author} {\bibfnamefont {S.~M.~H.}\ \bibnamefont
  {Rafsanjani}}, \ and\ \bibinfo {author} {\bibfnamefont {J.~H.}\ \bibnamefont
  {Eberly}},\ }\href {\doibase 10.1103/PhysRevA.85.043815} {\bibfield
  {journal} {\bibinfo  {journal} {Phys. Rev. A}\ }\textbf {\bibinfo {volume}
  {85}},\ \bibinfo {pages} {043815} (\bibinfo {year} {2012})}\BibitemShut
  {NoStop}%
\bibitem [{\citenamefont {Peng}\ \emph {et~al.}(2012)\citenamefont {Peng},
  \citenamefont {Ren}, \citenamefont {Guo},\ and\ \citenamefont {Ju}}]{Peng12}%
  \BibitemOpen
  \bibfield  {author} {\bibinfo {author} {\bibfnamefont {J.}~\bibnamefont
  {Peng}}, \bibinfo {author} {\bibfnamefont {Z.}~\bibnamefont {Ren}}, \bibinfo
  {author} {\bibfnamefont {G.}~\bibnamefont {Guo}}, \ and\ \bibinfo {author}
  {\bibfnamefont {G.}~\bibnamefont {Ju}},\ }\href {\doibase
  10.1088/1751-8113/45/36/365302} {\bibfield  {journal} {\bibinfo  {journal}
  {Journal of Physics A: Mathematical and Theoretical}\ }\textbf {\bibinfo
  {volume} {45}},\ \bibinfo {pages} {365302} (\bibinfo {year}
  {2012})}\BibitemShut {NoStop}%
\bibitem [{\citenamefont {Lee}\ and\ \citenamefont {Law}(2013)}]{Lee13}%
  \BibitemOpen
  \bibfield  {author} {\bibinfo {author} {\bibfnamefont {K.~M.~C.}\
  \bibnamefont {Lee}}\ and\ \bibinfo {author} {\bibfnamefont {C.~K.}\
  \bibnamefont {Law}},\ }\href {\doibase 10.1103/PhysRevA.88.015802} {\bibfield
   {journal} {\bibinfo  {journal} {Phys. Rev. A}\ }\textbf {\bibinfo {volume}
  {88}},\ \bibinfo {pages} {015802} (\bibinfo {year} {2013})}\BibitemShut
  {NoStop}%
\bibitem [{\citenamefont {Chilingaryan}\ and\ \citenamefont
  {Rodr{\'{\i}}guez-Lara}(2013)}]{Chilingaryan13}%
  \BibitemOpen
  \bibfield  {author} {\bibinfo {author} {\bibfnamefont {S.~A.}\ \bibnamefont
  {Chilingaryan}}\ and\ \bibinfo {author} {\bibfnamefont {B.~M.}\ \bibnamefont
  {Rodr{\'{\i}}guez-Lara}},\ }\href {\doibase 10.1088/1751-8113/46/33/335301}
  {\bibfield  {journal} {\bibinfo  {journal} {Journal of Physics A:
  Mathematical and Theoretical}\ }\textbf {\bibinfo {volume} {46}},\ \bibinfo
  {pages} {335301} (\bibinfo {year} {2013})}\BibitemShut {NoStop}%
\bibitem [{\citenamefont {Wang}\ \emph {et~al.}(2014)\citenamefont {Wang},
  \citenamefont {He}, \citenamefont {Duan}, \citenamefont {Zhao},\ and\
  \citenamefont {Chen}}]{Wang14}%
  \BibitemOpen
  \bibfield  {author} {\bibinfo {author} {\bibfnamefont {H.}~\bibnamefont
  {Wang}}, \bibinfo {author} {\bibfnamefont {S.}~\bibnamefont {He}}, \bibinfo
  {author} {\bibfnamefont {L.}~\bibnamefont {Duan}}, \bibinfo {author}
  {\bibfnamefont {Y.}~\bibnamefont {Zhao}}, \ and\ \bibinfo {author}
  {\bibfnamefont {Q.-H.}\ \bibnamefont {Chen}},\ }\href {\doibase
  10.1209/0295-5075/106/54001} {\bibfield  {journal} {\bibinfo  {journal}
  {{EPL} (Europhysics Letters)}\ }\textbf {\bibinfo {volume} {106}},\ \bibinfo
  {pages} {54001} (\bibinfo {year} {2014})}\BibitemShut {NoStop}%
\bibitem [{\citenamefont {Peng}\ \emph {et~al.}(2014)\citenamefont {Peng},
  \citenamefont {Ren}, \citenamefont {Braak}, \citenamefont {Guo},
  \citenamefont {Ju}, \citenamefont {Zhang},\ and\ \citenamefont
  {Guo}}]{Peng14}%
  \BibitemOpen
  \bibfield  {author} {\bibinfo {author} {\bibfnamefont {J.}~\bibnamefont
  {Peng}}, \bibinfo {author} {\bibfnamefont {Z.}~\bibnamefont {Ren}}, \bibinfo
  {author} {\bibfnamefont {D.}~\bibnamefont {Braak}}, \bibinfo {author}
  {\bibfnamefont {G.}~\bibnamefont {Guo}}, \bibinfo {author} {\bibfnamefont
  {G.}~\bibnamefont {Ju}}, \bibinfo {author} {\bibfnamefont {X.}~\bibnamefont
  {Zhang}}, \ and\ \bibinfo {author} {\bibfnamefont {X.}~\bibnamefont {Guo}},\
  }\href {\doibase 10.1088/1751-8113/47/26/265303} {\bibfield  {journal}
  {\bibinfo  {journal} {Journal of Physics A: Mathematical and Theoretical}\
  }\textbf {\bibinfo {volume} {47}},\ \bibinfo {pages} {265303} (\bibinfo
  {year} {2014})}\BibitemShut {NoStop}%
\bibitem [{\citenamefont {Peng}\ \emph {et~al.}(2021)\citenamefont {Peng},
  \citenamefont {Zheng}, \citenamefont {Yu}, \citenamefont {Tang},
  \citenamefont {Barrios}, \citenamefont {Zhong}, \citenamefont {Solano},
  \citenamefont {Albarr\'an-Arriagada},\ and\ \citenamefont {Lamata}}]{Peng21}%
  \BibitemOpen
  \bibfield  {author} {\bibinfo {author} {\bibfnamefont {J.}~\bibnamefont
  {Peng}}, \bibinfo {author} {\bibfnamefont {J.}~\bibnamefont {Zheng}},
  \bibinfo {author} {\bibfnamefont {J.}~\bibnamefont {Yu}}, \bibinfo {author}
  {\bibfnamefont {P.}~\bibnamefont {Tang}}, \bibinfo {author} {\bibfnamefont
  {G.~A.}\ \bibnamefont {Barrios}}, \bibinfo {author} {\bibfnamefont
  {J.}~\bibnamefont {Zhong}}, \bibinfo {author} {\bibfnamefont
  {E.}~\bibnamefont {Solano}}, \bibinfo {author} {\bibfnamefont
  {F.}~\bibnamefont {Albarr\'an-Arriagada}}, \ and\ \bibinfo {author}
  {\bibfnamefont {L.}~\bibnamefont {Lamata}},\ }\href {\doibase
  10.1103/PhysRevLett.127.043604} {\bibfield  {journal} {\bibinfo  {journal}
  {Phys. Rev. Lett.}\ }\textbf {\bibinfo {volume} {127}},\ \bibinfo {pages}
  {043604} (\bibinfo {year} {2021})}\BibitemShut {NoStop}%
\bibitem [{\citenamefont {Zhang}\ \emph
  {et~al.}(2021{\natexlab{a}})\citenamefont {Zhang}, \citenamefont {Hu},
  \citenamefont {Fu}, \citenamefont {Luo}, \citenamefont {Pu},\ and\
  \citenamefont {Zhang}}]{Zhang21prl}%
  \BibitemOpen
  \bibfield  {author} {\bibinfo {author} {\bibfnamefont {Y.-Y.}\ \bibnamefont
  {Zhang}}, \bibinfo {author} {\bibfnamefont {Z.-X.}\ \bibnamefont {Hu}},
  \bibinfo {author} {\bibfnamefont {L.}~\bibnamefont {Fu}}, \bibinfo {author}
  {\bibfnamefont {H.-G.}\ \bibnamefont {Luo}}, \bibinfo {author} {\bibfnamefont
  {H.}~\bibnamefont {Pu}}, \ and\ \bibinfo {author} {\bibfnamefont {X.-F.}\
  \bibnamefont {Zhang}},\ }\href {\doibase 10.1103/PhysRevLett.127.063602}
  {\bibfield  {journal} {\bibinfo  {journal} {Phys. Rev. Lett.}\ }\textbf
  {\bibinfo {volume} {127}},\ \bibinfo {pages} {063602} (\bibinfo {year}
  {2021}{\natexlab{a}})}\BibitemShut {NoStop}%
\bibitem [{\citenamefont {Nataf}\ and\ \citenamefont {Ciuti}(2011)}]{Nataf11}%
  \BibitemOpen
  \bibfield  {author} {\bibinfo {author} {\bibfnamefont {P.}~\bibnamefont
  {Nataf}}\ and\ \bibinfo {author} {\bibfnamefont {C.}~\bibnamefont {Ciuti}},\
  }\href {\doibase 10.1103/PhysRevLett.107.190402} {\bibfield  {journal}
  {\bibinfo  {journal} {Phys. Rev. Lett.}\ }\textbf {\bibinfo {volume} {107}},\
  \bibinfo {pages} {190402} (\bibinfo {year} {2011})}\BibitemShut {NoStop}%
\bibitem [{\citenamefont {Lizuain}\ \emph {et~al.}(2010)\citenamefont
  {Lizuain}, \citenamefont {Casanova}, \citenamefont {Garc\'{\i}a-Ripoll},
  \citenamefont {Muga},\ and\ \citenamefont {Solano}}]{Lizuain10}%
  \BibitemOpen
  \bibfield  {author} {\bibinfo {author} {\bibfnamefont {I.}~\bibnamefont
  {Lizuain}}, \bibinfo {author} {\bibfnamefont {J.}~\bibnamefont {Casanova}},
  \bibinfo {author} {\bibfnamefont {J.~J.}\ \bibnamefont {Garc\'{\i}a-Ripoll}},
  \bibinfo {author} {\bibfnamefont {J.~G.}\ \bibnamefont {Muga}}, \ and\
  \bibinfo {author} {\bibfnamefont {E.}~\bibnamefont {Solano}},\ }\href
  {\doibase 10.1103/PhysRevA.81.062131} {\bibfield  {journal} {\bibinfo
  {journal} {Phys. Rev. A}\ }\textbf {\bibinfo {volume} {81}},\ \bibinfo
  {pages} {062131} (\bibinfo {year} {2010})}\BibitemShut {NoStop}%
\bibitem [{\citenamefont {Carusotto}\ and\ \citenamefont
  {Ciuti}(2013)}]{Crusotto13}%
  \BibitemOpen
  \bibfield  {author} {\bibinfo {author} {\bibfnamefont {I.}~\bibnamefont
  {Carusotto}}\ and\ \bibinfo {author} {\bibfnamefont {C.}~\bibnamefont
  {Ciuti}},\ }\href {\doibase 10.1103/RevModPhys.85.299} {\bibfield  {journal}
  {\bibinfo  {journal} {Rev. Mod. Phys.}\ }\textbf {\bibinfo {volume} {85}},\
  \bibinfo {pages} {299} (\bibinfo {year} {2013})}\BibitemShut {NoStop}%
\bibitem [{\citenamefont {Anappara}\ \emph {et~al.}(2009)\citenamefont
  {Anappara}, \citenamefont {De~Liberato}, \citenamefont {Tredicucci},
  \citenamefont {Ciuti}, \citenamefont {Biasiol}, \citenamefont {Sorba},\ and\
  \citenamefont {Beltram}}]{Anappara09}%
  \BibitemOpen
  \bibfield  {author} {\bibinfo {author} {\bibfnamefont {A.~A.}\ \bibnamefont
  {Anappara}}, \bibinfo {author} {\bibfnamefont {S.}~\bibnamefont
  {De~Liberato}}, \bibinfo {author} {\bibfnamefont {A.}~\bibnamefont
  {Tredicucci}}, \bibinfo {author} {\bibfnamefont {C.}~\bibnamefont {Ciuti}},
  \bibinfo {author} {\bibfnamefont {G.}~\bibnamefont {Biasiol}}, \bibinfo
  {author} {\bibfnamefont {L.}~\bibnamefont {Sorba}}, \ and\ \bibinfo {author}
  {\bibfnamefont {F.}~\bibnamefont {Beltram}},\ }\href {\doibase
  10.1103/PhysRevB.79.201303} {\bibfield  {journal} {\bibinfo  {journal} {Phys.
  Rev. B}\ }\textbf {\bibinfo {volume} {79}},\ \bibinfo {pages} {201303}
  (\bibinfo {year} {2009})}\BibitemShut {NoStop}%
\bibitem [{\citenamefont {Todorov}\ \emph {et~al.}(2010)\citenamefont
  {Todorov}, \citenamefont {Andrews}, \citenamefont {Colombelli}, \citenamefont
  {De~Liberato}, \citenamefont {Ciuti}, \citenamefont {Klang}, \citenamefont
  {Strasser},\ and\ \citenamefont {Sirtori}}]{Todorov10}%
  \BibitemOpen
  \bibfield  {author} {\bibinfo {author} {\bibfnamefont {Y.}~\bibnamefont
  {Todorov}}, \bibinfo {author} {\bibfnamefont {A.~M.}\ \bibnamefont
  {Andrews}}, \bibinfo {author} {\bibfnamefont {R.}~\bibnamefont {Colombelli}},
  \bibinfo {author} {\bibfnamefont {S.}~\bibnamefont {De~Liberato}}, \bibinfo
  {author} {\bibfnamefont {C.}~\bibnamefont {Ciuti}}, \bibinfo {author}
  {\bibfnamefont {P.}~\bibnamefont {Klang}}, \bibinfo {author} {\bibfnamefont
  {G.}~\bibnamefont {Strasser}}, \ and\ \bibinfo {author} {\bibfnamefont
  {C.}~\bibnamefont {Sirtori}},\ }\href {\doibase
  10.1103/PhysRevLett.105.196402} {\bibfield  {journal} {\bibinfo  {journal}
  {Phys. Rev. Lett.}\ }\textbf {\bibinfo {volume} {105}},\ \bibinfo {pages}
  {196402} (\bibinfo {year} {2010})}\BibitemShut {NoStop}%
\bibitem [{\citenamefont {Zhang}\ and\ \citenamefont {Chen}(2015)}]{Zhang15}%
  \BibitemOpen
  \bibfield  {author} {\bibinfo {author} {\bibfnamefont {Y.-Y.}\ \bibnamefont
  {Zhang}}\ and\ \bibinfo {author} {\bibfnamefont {Q.-H.}\ \bibnamefont
  {Chen}},\ }\href {\doibase 10.1103/PhysRevA.91.013814} {\bibfield  {journal}
  {\bibinfo  {journal} {Phys. Rev. A}\ }\textbf {\bibinfo {volume} {91}},\
  \bibinfo {pages} {013814} (\bibinfo {year} {2015})}\BibitemShut {NoStop}%
\bibitem [{\citenamefont {Duan}\ \emph {et~al.}(2015)\citenamefont {Duan},
  \citenamefont {He},\ and\ \citenamefont {Chen}}]{Duan15}%
  \BibitemOpen
  \bibfield  {author} {\bibinfo {author} {\bibfnamefont {L.}~\bibnamefont
  {Duan}}, \bibinfo {author} {\bibfnamefont {S.}~\bibnamefont {He}}, \ and\
  \bibinfo {author} {\bibfnamefont {Q.-H.}\ \bibnamefont {Chen}},\ }\href
  {\doibase https://doi.org/10.1016/j.aop.2015.02.003} {\bibfield  {journal}
  {\bibinfo  {journal} {Annals of Physics}\ }\textbf {\bibinfo {volume}
  {355}},\ \bibinfo {pages} {121} (\bibinfo {year} {2015})}\BibitemShut
  {NoStop}%
\bibitem [{\citenamefont {Dong}(2016)}]{Dong16}%
  \BibitemOpen
  \bibfield  {author} {\bibinfo {author} {\bibfnamefont {K.}~\bibnamefont
  {Dong}},\ }\href {\doibase 10.1088/1674-1056/25/12/124202} {\bibfield
  {journal} {\bibinfo  {journal} {Chinese Physics B}\ }\textbf {\bibinfo
  {volume} {25}},\ \bibinfo {pages} {124202} (\bibinfo {year}
  {2016})}\BibitemShut {NoStop}%
\bibitem [{\citenamefont {Mao}\ \emph {et~al.}(2019)\citenamefont {Mao},
  \citenamefont {Li}, \citenamefont {Wang}, \citenamefont {You}, \citenamefont
  {Wu}, \citenamefont {Liu},\ and\ \citenamefont {Luo}}]{Mao19}%
  \BibitemOpen
  \bibfield  {author} {\bibinfo {author} {\bibfnamefont {B.-B.}\ \bibnamefont
  {Mao}}, \bibinfo {author} {\bibfnamefont {L.}~\bibnamefont {Li}}, \bibinfo
  {author} {\bibfnamefont {Y.}~\bibnamefont {Wang}}, \bibinfo {author}
  {\bibfnamefont {W.-L.}\ \bibnamefont {You}}, \bibinfo {author} {\bibfnamefont
  {W.}~\bibnamefont {Wu}}, \bibinfo {author} {\bibfnamefont {M.}~\bibnamefont
  {Liu}}, \ and\ \bibinfo {author} {\bibfnamefont {H.-G.}\ \bibnamefont
  {Luo}},\ }\href {\doibase 10.1103/PhysRevA.99.033834} {\bibfield  {journal}
  {\bibinfo  {journal} {Phys. Rev. A}\ }\textbf {\bibinfo {volume} {99}},\
  \bibinfo {pages} {033834} (\bibinfo {year} {2019})}\BibitemShut {NoStop}%
\bibitem [{\citenamefont {Sun}\ \emph {et~al.}(2020)\citenamefont {Sun},
  \citenamefont {Cong}, \citenamefont {Eckle}, \citenamefont {Ying},\ and\
  \citenamefont {Luo}}]{Sun20}%
  \BibitemOpen
  \bibfield  {author} {\bibinfo {author} {\bibfnamefont {X.-M.}\ \bibnamefont
  {Sun}}, \bibinfo {author} {\bibfnamefont {L.}~\bibnamefont {Cong}}, \bibinfo
  {author} {\bibfnamefont {H.-P.}\ \bibnamefont {Eckle}}, \bibinfo {author}
  {\bibfnamefont {Z.-J.}\ \bibnamefont {Ying}}, \ and\ \bibinfo {author}
  {\bibfnamefont {H.-G.}\ \bibnamefont {Luo}},\ }\href {\doibase
  10.1103/PhysRevA.101.063832} {\bibfield  {journal} {\bibinfo  {journal}
  {Phys. Rev. A}\ }\textbf {\bibinfo {volume} {101}},\ \bibinfo {pages}
  {063832} (\bibinfo {year} {2020})}\BibitemShut {NoStop}%
\bibitem [{\citenamefont {Yan}\ \emph {et~al.}(2021)\citenamefont {Yan},
  \citenamefont {Qu}, \citenamefont {Xu}, \citenamefont {Zhang},\ and\
  \citenamefont {Ma}}]{Yan21}%
  \BibitemOpen
  \bibfield  {author} {\bibinfo {author} {\bibfnamefont {Z.}~\bibnamefont
  {Yan}}, \bibinfo {author} {\bibfnamefont {P.}~\bibnamefont {Qu}}, \bibinfo
  {author} {\bibfnamefont {B.}~\bibnamefont {Xu}}, \bibinfo {author}
  {\bibfnamefont {S.}~\bibnamefont {Zhang}}, \ and\ \bibinfo {author}
  {\bibfnamefont {J.}~\bibnamefont {Ma}},\ }\href {\doibase
  10.1142/S0217984921502134} {\bibfield  {journal} {\bibinfo  {journal} {Modern
  Physics Letters B}\ }\textbf {\bibinfo {volume} {35}},\ \bibinfo {pages}
  {2150213} (\bibinfo {year} {2021})}\BibitemShut {NoStop}%
\bibitem [{\citenamefont {Zhang}\ \emph
  {et~al.}(2021{\natexlab{b}})\citenamefont {Zhang}, \citenamefont {Han},\ and\
  \citenamefont {Chen}}]{Zhang21}%
  \BibitemOpen
  \bibfield  {author} {\bibinfo {author} {\bibfnamefont {Y.-L.}\ \bibnamefont
  {Zhang}}, \bibinfo {author} {\bibfnamefont {R.-S.}\ \bibnamefont {Han}}, \
  and\ \bibinfo {author} {\bibfnamefont {L.}~\bibnamefont {Chen}},\ }\href@noop
  {} {\bibfield  {journal} {\bibinfo  {journal} {Int. J. Theor. Phys.}\
  }\textbf {\bibinfo {volume} {60}},\ \bibinfo {pages} {1384} (\bibinfo {year}
  {2021}{\natexlab{b}})}\BibitemShut {NoStop}%
\bibitem [{\citenamefont {Liu}\ \emph {et~al.}(2021)\citenamefont {Liu},
  \citenamefont {Liu}, \citenamefont {Ying},\ and\ \citenamefont
  {Luo}}]{Liu21}%
  \BibitemOpen
  \bibfield  {author} {\bibinfo {author} {\bibfnamefont {J.}~\bibnamefont
  {Liu}}, \bibinfo {author} {\bibfnamefont {M.}~\bibnamefont {Liu}}, \bibinfo
  {author} {\bibfnamefont {Z.-J.}\ \bibnamefont {Ying}}, \ and\ \bibinfo
  {author} {\bibfnamefont {H.-G.}\ \bibnamefont {Luo}},\ }\href {\doibase
  https://doi.org/10.1002/qute.202000139} {\bibfield  {journal} {\bibinfo
  {journal} {Advanced Quantum Technologies}\ }\textbf {\bibinfo {volume} {4}},\
  \bibinfo {pages} {2000139} (\bibinfo {year} {2021})}\BibitemShut {NoStop}%
\bibitem [{\citenamefont {Mao}\ \emph {et~al.}(2021)\citenamefont {Mao},
  \citenamefont {Li}, \citenamefont {You},\ and\ \citenamefont {Liu}}]{Mao21}%
  \BibitemOpen
  \bibfield  {author} {\bibinfo {author} {\bibfnamefont {B.-B.}\ \bibnamefont
  {Mao}}, \bibinfo {author} {\bibfnamefont {L.}~\bibnamefont {Li}}, \bibinfo
  {author} {\bibfnamefont {W.-L.}\ \bibnamefont {You}}, \ and\ \bibinfo
  {author} {\bibfnamefont {M.}~\bibnamefont {Liu}},\ }\href {\doibase
  https://doi.org/10.1016/j.physa.2020.125534} {\bibfield  {journal} {\bibinfo
  {journal} {Physica A: Statistical Mechanics and its Applications}\ }\textbf
  {\bibinfo {volume} {564}},\ \bibinfo {pages} {125534} (\bibinfo {year}
  {2021})}\BibitemShut {NoStop}%
\bibitem [{\citenamefont {Mao}\ \emph {et~al.}(2015)\citenamefont {Mao},
  \citenamefont {Huai},\ and\ \citenamefont {Zhang}}]{Mao15}%
  \BibitemOpen
  \bibfield  {author} {\bibinfo {author} {\bibfnamefont {L.}~\bibnamefont
  {Mao}}, \bibinfo {author} {\bibfnamefont {S.}~\bibnamefont {Huai}}, \ and\
  \bibinfo {author} {\bibfnamefont {Y.}~\bibnamefont {Zhang}},\ }\href
  {\doibase 10.1088/1751-8113/48/34/345302} {\bibfield  {journal} {\bibinfo
  {journal} {Journal of Physics A: Mathematical and Theoretical}\ }\textbf
  {\bibinfo {volume} {48}},\ \bibinfo {pages} {345302} (\bibinfo {year}
  {2015})}\BibitemShut {NoStop}%
\bibitem [{\citenamefont {Grimaudo}\ \emph
  {et~al.}(2019{\natexlab{a}})\citenamefont {Grimaudo}, \citenamefont
  {Vitanov},\ and\ \citenamefont {Messina}}]{GVM1}%
  \BibitemOpen
  \bibfield  {author} {\bibinfo {author} {\bibfnamefont {R.}~\bibnamefont
  {Grimaudo}}, \bibinfo {author} {\bibfnamefont {N.~V.}\ \bibnamefont
  {Vitanov}}, \ and\ \bibinfo {author} {\bibfnamefont {A.}~\bibnamefont
  {Messina}},\ }\href {\doibase 10.1103/PhysRevB.99.174416} {\bibfield
  {journal} {\bibinfo  {journal} {Phys. Rev. B}\ }\textbf {\bibinfo {volume}
  {99}},\ \bibinfo {pages} {174416} (\bibinfo {year}
  {2019}{\natexlab{a}})}\BibitemShut {NoStop}%
\bibitem [{\citenamefont {Grimaudo}\ \emph
  {et~al.}(2022{\natexlab{a}})\citenamefont {Grimaudo}, \citenamefont
  {Messina}, \citenamefont {Nakazato}, \citenamefont {Sergi},\ and\
  \citenamefont {Valenti}}]{GMNSV}%
  \BibitemOpen
  \bibfield  {author} {\bibinfo {author} {\bibfnamefont {R.}~\bibnamefont
  {Grimaudo}}, \bibinfo {author} {\bibfnamefont {A.}~\bibnamefont {Messina}},
  \bibinfo {author} {\bibfnamefont {H.}~\bibnamefont {Nakazato}}, \bibinfo
  {author} {\bibfnamefont {A.}~\bibnamefont {Sergi}}, \ and\ \bibinfo {author}
  {\bibfnamefont {D.}~\bibnamefont {Valenti}},\ }\href@noop {} {\bibfield
  {journal} {\bibinfo  {journal} {arXiv preprint arXiv:2205.09367}\ } (\bibinfo
  {year} {2022}{\natexlab{a}})}\BibitemShut {NoStop}%
\bibitem [{Note1()}]{Note1}%
  \BibitemOpen
  \bibinfo {note} {The original QRM is obtained by putting the longitudinal
  ($\protect \hat {z}$) field equal to zero. Moreover, a $\pi /2$-rotation
  around the $\protect \hat {y}$-axis has to be performed for both $H_a$ and
  $H_b$ to get the standard form of the asymmetric QRM.}\BibitemShut {Stop}%
\bibitem [{\citenamefont {Frisk~Kockum}\ \emph {et~al.}(2019)\citenamefont
  {Frisk~Kockum}, \citenamefont {Miranowicz}, \citenamefont {De~Liberato},
  \citenamefont {Savasta},\ and\ \citenamefont {Nori}}]{Kockum19}%
  \BibitemOpen
  \bibfield  {author} {\bibinfo {author} {\bibfnamefont {A.}~\bibnamefont
  {Frisk~Kockum}}, \bibinfo {author} {\bibfnamefont {A.}~\bibnamefont
  {Miranowicz}}, \bibinfo {author} {\bibfnamefont {S.}~\bibnamefont
  {De~Liberato}}, \bibinfo {author} {\bibfnamefont {S.}~\bibnamefont
  {Savasta}}, \ and\ \bibinfo {author} {\bibfnamefont {F.}~\bibnamefont
  {Nori}},\ }\href@noop {} {\bibfield  {journal} {\bibinfo  {journal} {Nature
  Rev. Phys.}\ }\textbf {\bibinfo {volume} {1}},\ \bibinfo {pages} {19}
  (\bibinfo {year} {2019})}\BibitemShut {NoStop}%
\bibitem [{\citenamefont {Xie}\ \emph {et~al.}(2017)\citenamefont {Xie},
  \citenamefont {Zhong}, \citenamefont {Batchelor},\ and\ \citenamefont
  {Lee}}]{Xie17}%
  \BibitemOpen
  \bibfield  {author} {\bibinfo {author} {\bibfnamefont {Q.}~\bibnamefont
  {Xie}}, \bibinfo {author} {\bibfnamefont {H.}~\bibnamefont {Zhong}}, \bibinfo
  {author} {\bibfnamefont {M.~T.}\ \bibnamefont {Batchelor}}, \ and\ \bibinfo
  {author} {\bibfnamefont {C.}~\bibnamefont {Lee}},\ }\href {\doibase
  10.1088/1751-8121/aa5a65} {\bibfield  {journal} {\bibinfo  {journal} {Journal
  of Physics A: Mathematical and Theoretical}\ }\textbf {\bibinfo {volume}
  {50}},\ \bibinfo {pages} {113001} (\bibinfo {year} {2017})}\BibitemShut
  {NoStop}%
\bibitem [{\citenamefont {Cahill}\ and\ \citenamefont
  {Glauber}(1969)}]{Glauber69}%
  \BibitemOpen
  \bibfield  {author} {\bibinfo {author} {\bibfnamefont {K.~E.}\ \bibnamefont
  {Cahill}}\ and\ \bibinfo {author} {\bibfnamefont {R.~J.}\ \bibnamefont
  {Glauber}},\ }\href {\doibase 10.1103/PhysRev.177.1857} {\bibfield  {journal}
  {\bibinfo  {journal} {Phys. Rev.}\ }\textbf {\bibinfo {volume} {177}},\
  \bibinfo {pages} {1857} (\bibinfo {year} {1969})}\BibitemShut {NoStop}%
\bibitem [{\citenamefont {Zhong}\ \emph {et~al.}(2013)\citenamefont {Zhong},
  \citenamefont {Xie}, \citenamefont {Batchelor},\ and\ \citenamefont
  {Lee}}]{Zhong13}%
  \BibitemOpen
  \bibfield  {author} {\bibinfo {author} {\bibfnamefont {H.}~\bibnamefont
  {Zhong}}, \bibinfo {author} {\bibfnamefont {Q.}~\bibnamefont {Xie}}, \bibinfo
  {author} {\bibfnamefont {M.~T.}\ \bibnamefont {Batchelor}}, \ and\ \bibinfo
  {author} {\bibfnamefont {C.}~\bibnamefont {Lee}},\ }\href {\doibase
  10.1088/1751-8113/46/41/415302} {\bibfield  {journal} {\bibinfo  {journal}
  {Journal of Physics A: Mathematical and Theoretical}\ }\textbf {\bibinfo
  {volume} {46}},\ \bibinfo {pages} {415302} (\bibinfo {year}
  {2013})}\BibitemShut {NoStop}%
\bibitem [{Note2()}]{Note2}%
  \BibitemOpen
  \bibinfo {note} {See Eqs. 32, 33 and 34 in Ref. \cite {Zhong13}; our
  eigenstate looks different from the one in Ref. \cite {Zhong13} since our
  effective Hamiltonian $H_a$ (and $H_b$) is a QRM Hamiltonian rotated of $\pi
  /2$ around the $\protect \hat {y}$-axis}\BibitemShut {NoStop}%
\bibitem [{\citenamefont {Wootters}(1998)}]{Wootters98}%
  \BibitemOpen
  \bibfield  {author} {\bibinfo {author} {\bibfnamefont {W.~K.}\ \bibnamefont
  {Wootters}},\ }\href {\doibase 10.1103/PhysRevLett.80.2245} {\bibfield
  {journal} {\bibinfo  {journal} {Phys. Rev. Lett.}\ }\textbf {\bibinfo
  {volume} {80}},\ \bibinfo {pages} {2245} (\bibinfo {year}
  {1998})}\BibitemShut {NoStop}%
\bibitem [{\citenamefont {Vandersypen}\ and\ \citenamefont
  {Chuang}(2005)}]{Vandersypen05}%
  \BibitemOpen
  \bibfield  {author} {\bibinfo {author} {\bibfnamefont {L.~M.~K.}\
  \bibnamefont {Vandersypen}}\ and\ \bibinfo {author} {\bibfnamefont {I.~L.}\
  \bibnamefont {Chuang}},\ }\href {\doibase 10.1103/RevModPhys.76.1037}
  {\bibfield  {journal} {\bibinfo  {journal} {Rev. Mod. Phys.}\ }\textbf
  {\bibinfo {volume} {76}},\ \bibinfo {pages} {1037} (\bibinfo {year}
  {2005})}\BibitemShut {NoStop}%
\bibitem [{\citenamefont {Weidt}\ \emph {et~al.}(2016)\citenamefont {Weidt},
  \citenamefont {Randall}, \citenamefont {Webster}, \citenamefont {Lake},
  \citenamefont {Webb}, \citenamefont {Cohen}, \citenamefont {Navickas},
  \citenamefont {Lekitsch}, \citenamefont {Retzker},\ and\ \citenamefont
  {Hensinger}}]{Weidt16}%
  \BibitemOpen
  \bibfield  {author} {\bibinfo {author} {\bibfnamefont {S.}~\bibnamefont
  {Weidt}}, \bibinfo {author} {\bibfnamefont {J.}~\bibnamefont {Randall}},
  \bibinfo {author} {\bibfnamefont {S.~C.}\ \bibnamefont {Webster}}, \bibinfo
  {author} {\bibfnamefont {K.}~\bibnamefont {Lake}}, \bibinfo {author}
  {\bibfnamefont {A.~E.}\ \bibnamefont {Webb}}, \bibinfo {author}
  {\bibfnamefont {I.}~\bibnamefont {Cohen}}, \bibinfo {author} {\bibfnamefont
  {T.}~\bibnamefont {Navickas}}, \bibinfo {author} {\bibfnamefont
  {B.}~\bibnamefont {Lekitsch}}, \bibinfo {author} {\bibfnamefont
  {A.}~\bibnamefont {Retzker}}, \ and\ \bibinfo {author} {\bibfnamefont
  {W.~K.}\ \bibnamefont {Hensinger}},\ }\href {\doibase
  10.1103/PhysRevLett.117.220501} {\bibfield  {journal} {\bibinfo  {journal}
  {Phys. Rev. Lett.}\ }\textbf {\bibinfo {volume} {117}},\ \bibinfo {pages}
  {220501} (\bibinfo {year} {2016})}\BibitemShut {NoStop}%
\bibitem [{\citenamefont {Gaetan}\ \emph {et~al.}(2009)\citenamefont {Gaetan},
  \citenamefont {Miroshnychenko}, \citenamefont {Wilk}, \citenamefont {Chotia},
  \citenamefont {Viteau}, \citenamefont {Comparat}, \citenamefont {Pillet},
  \citenamefont {Browaeys},\ and\ \citenamefont {Grangier}}]{Gaetan09}%
  \BibitemOpen
  \bibfield  {author} {\bibinfo {author} {\bibfnamefont {A.}~\bibnamefont
  {Gaetan}}, \bibinfo {author} {\bibfnamefont {Y.}~\bibnamefont
  {Miroshnychenko}}, \bibinfo {author} {\bibfnamefont {T.}~\bibnamefont
  {Wilk}}, \bibinfo {author} {\bibfnamefont {A.}~\bibnamefont {Chotia}},
  \bibinfo {author} {\bibfnamefont {M.}~\bibnamefont {Viteau}}, \bibinfo
  {author} {\bibfnamefont {D.}~\bibnamefont {Comparat}}, \bibinfo {author}
  {\bibfnamefont {P.}~\bibnamefont {Pillet}}, \bibinfo {author} {\bibfnamefont
  {A.}~\bibnamefont {Browaeys}}, \ and\ \bibinfo {author} {\bibfnamefont
  {P.}~\bibnamefont {Grangier}},\ }\href@noop {} {\bibfield  {journal}
  {\bibinfo  {journal} {Nat. Phys.}\ }\textbf {\bibinfo {volume} {5}},\
  \bibinfo {pages} {115} (\bibinfo {year} {2009})}\BibitemShut {NoStop}%
\bibitem [{\citenamefont {Urban}\ \emph {et~al.}(2009)\citenamefont {Urban},
  \citenamefont {Johnson}, \citenamefont {Henage}, \citenamefont {Isenhower},
  \citenamefont {Yavuz}, \citenamefont {Walker},\ and\ \citenamefont
  {Saffman}}]{Urban09}%
  \BibitemOpen
  \bibfield  {author} {\bibinfo {author} {\bibfnamefont {E.}~\bibnamefont
  {Urban}}, \bibinfo {author} {\bibfnamefont {T.~A.}\ \bibnamefont {Johnson}},
  \bibinfo {author} {\bibfnamefont {T.}~\bibnamefont {Henage}}, \bibinfo
  {author} {\bibfnamefont {L.}~\bibnamefont {Isenhower}}, \bibinfo {author}
  {\bibfnamefont {D.}~\bibnamefont {Yavuz}}, \bibinfo {author} {\bibfnamefont
  {T.}~\bibnamefont {Walker}}, \ and\ \bibinfo {author} {\bibfnamefont
  {M.}~\bibnamefont {Saffman}},\ }\href@noop {} {\bibfield  {journal} {\bibinfo
   {journal} {Nat. Phys.}\ }\textbf {\bibinfo {volume} {5}},\ \bibinfo {pages}
  {110} (\bibinfo {year} {2009})}\BibitemShut {NoStop}%
\bibitem [{\citenamefont {Grimaudo}\ \emph {et~al.}(2017)\citenamefont
  {Grimaudo}, \citenamefont {Messina}, \citenamefont {Ivanov},\ and\
  \citenamefont {Vitanov}}]{GMIV}%
  \BibitemOpen
  \bibfield  {author} {\bibinfo {author} {\bibfnamefont {R.}~\bibnamefont
  {Grimaudo}}, \bibinfo {author} {\bibfnamefont {A.}~\bibnamefont {Messina}},
  \bibinfo {author} {\bibfnamefont {P.~A.}\ \bibnamefont {Ivanov}}, \ and\
  \bibinfo {author} {\bibfnamefont {N.~V.}\ \bibnamefont {Vitanov}},\ }\href
  {\doibase 10.1088/1751-8121/aa5fb6} {\bibfield  {journal} {\bibinfo
  {journal} {J. Phys. A Math. Theor.}\ }\textbf {\bibinfo {volume} {50}},\
  \bibinfo {pages} {175301} (\bibinfo {year} {2017})}\BibitemShut {NoStop}%
\bibitem [{\citenamefont {Grimaudo}\ \emph
  {et~al.}(2019{\natexlab{b}})\citenamefont {Grimaudo}, \citenamefont
  {Vitanov},\ and\ \citenamefont {Messina}}]{GMV2}%
  \BibitemOpen
  \bibfield  {author} {\bibinfo {author} {\bibfnamefont {R.}~\bibnamefont
  {Grimaudo}}, \bibinfo {author} {\bibfnamefont {N.~V.}\ \bibnamefont
  {Vitanov}}, \ and\ \bibinfo {author} {\bibfnamefont {A.}~\bibnamefont
  {Messina}},\ }\href {\doibase 10.1103/PhysRevB.99.214406} {\bibfield
  {journal} {\bibinfo  {journal} {Phys. Rev. B}\ }\textbf {\bibinfo {volume}
  {99}},\ \bibinfo {pages} {214406} (\bibinfo {year}
  {2019}{\natexlab{b}})}\BibitemShut {NoStop}%
\bibitem [{\citenamefont {Grimaudo}\ \emph
  {et~al.}(2019{\natexlab{c}})\citenamefont {Grimaudo}, \citenamefont {Man'ko},
  \citenamefont {Man'ko},\ and\ \citenamefont {Messina}}]{GMMM}%
  \BibitemOpen
  \bibfield  {author} {\bibinfo {author} {\bibfnamefont {R.}~\bibnamefont
  {Grimaudo}}, \bibinfo {author} {\bibfnamefont {V.~I.}\ \bibnamefont
  {Man'ko}}, \bibinfo {author} {\bibfnamefont {M.~A.}\ \bibnamefont {Man'ko}},
  \ and\ \bibinfo {author} {\bibfnamefont {A.}~\bibnamefont {Messina}},\ }\href
  {\doibase 10.1088/1402-4896/ab4305} {\bibfield  {journal} {\bibinfo
  {journal} {Phys. Scr.}\ }\textbf {\bibinfo {volume} {95}},\ \bibinfo {pages}
  {024004} (\bibinfo {year} {2019}{\natexlab{c}})}\BibitemShut {NoStop}%
\bibitem [{\citenamefont {Grimaudo}\ \emph
  {et~al.}(2018{\natexlab{a}})\citenamefont {Grimaudo}, \citenamefont {Lamata},
  \citenamefont {Solano},\ and\ \citenamefont {Messina}}]{GLSM}%
  \BibitemOpen
  \bibfield  {author} {\bibinfo {author} {\bibfnamefont {R.}~\bibnamefont
  {Grimaudo}}, \bibinfo {author} {\bibfnamefont {L.}~\bibnamefont {Lamata}},
  \bibinfo {author} {\bibfnamefont {E.}~\bibnamefont {Solano}}, \ and\ \bibinfo
  {author} {\bibfnamefont {A.}~\bibnamefont {Messina}},\ }\href {\doibase
  10.1103/PhysRevA.98.042330} {\bibfield  {journal} {\bibinfo  {journal} {Phys.
  Rev. A}\ }\textbf {\bibinfo {volume} {98}},\ \bibinfo {pages} {042330}
  (\bibinfo {year} {2018}{\natexlab{a}})}\BibitemShut {NoStop}%
\bibitem [{\citenamefont {Grimaudo}\ \emph
  {et~al.}(2018{\natexlab{b}})\citenamefont {Grimaudo}, \citenamefont
  {Belousov}, \citenamefont {Nakazato},\ and\ \citenamefont {Messina}}]{GBNM}%
  \BibitemOpen
  \bibfield  {author} {\bibinfo {author} {\bibfnamefont {R.}~\bibnamefont
  {Grimaudo}}, \bibinfo {author} {\bibfnamefont {Y.}~\bibnamefont {Belousov}},
  \bibinfo {author} {\bibfnamefont {H.}~\bibnamefont {Nakazato}}, \ and\
  \bibinfo {author} {\bibfnamefont {A.}~\bibnamefont {Messina}},\ }\href
  {\doibase https://doi.org/10.1016/j.aop.2018.03.012} {\bibfield  {journal}
  {\bibinfo  {journal} {Ann. Phys. (NY)}\ }\textbf {\bibinfo {volume} {392}},\
  \bibinfo {pages} {242} (\bibinfo {year} {2018}{\natexlab{b}})}\BibitemShut
  {NoStop}%
\bibitem [{\citenamefont {Grimaudo}\ \emph
  {et~al.}(2022{\natexlab{b}})\citenamefont {Grimaudo}, \citenamefont
  {Vitanov}, \citenamefont {Magalhães~de Castro}, \citenamefont {Valenti},\
  and\ \citenamefont {Messina}}]{GVdCVM}%
  \BibitemOpen
  \bibfield  {author} {\bibinfo {author} {\bibfnamefont {R.}~\bibnamefont
  {Grimaudo}}, \bibinfo {author} {\bibfnamefont {N.~V.}\ \bibnamefont
  {Vitanov}}, \bibinfo {author} {\bibfnamefont {A.~S.}\ \bibnamefont
  {Magalhães~de Castro}}, \bibinfo {author} {\bibfnamefont {D.}~\bibnamefont
  {Valenti}}, \ and\ \bibinfo {author} {\bibfnamefont {A.}~\bibnamefont
  {Messina}},\ }\href {\doibase https://doi.org/10.1002/prop.202200010}
  {\bibfield  {journal} {\bibinfo  {journal} {Fortschritte der Physik}\
  }\textbf {\bibinfo {volume} {70}},\ \bibinfo {pages} {2200010} (\bibinfo
  {year} {2022}{\natexlab{b}})}\BibitemShut {NoStop}%
\bibitem [{\citenamefont {Grimaudo}\ \emph
  {et~al.}(2022{\natexlab{c}})\citenamefont {Grimaudo}, \citenamefont
  {Magalhães~de Castro}, \citenamefont {Messina},\ and\ \citenamefont
  {Valenti}}]{GdCMV}%
  \BibitemOpen
  \bibfield  {author} {\bibinfo {author} {\bibfnamefont {R.}~\bibnamefont
  {Grimaudo}}, \bibinfo {author} {\bibfnamefont {A.~S.}\ \bibnamefont
  {Magalhães~de Castro}}, \bibinfo {author} {\bibfnamefont {A.}~\bibnamefont
  {Messina}}, \ and\ \bibinfo {author} {\bibfnamefont {D.}~\bibnamefont
  {Valenti}},\ }\href {\doibase https://doi.org/10.1002/prop.202200042}
  {\bibfield  {journal} {\bibinfo  {journal} {Fortschritte der Physik}\
  }\textbf {\bibinfo {volume} {70}},\ \bibinfo {pages} {2200042} (\bibinfo
  {year} {2022}{\natexlab{c}})}\BibitemShut {NoStop}%
\end{thebibliography}%

\end{document}